\documentclass[times,preprint]{aastex631}

\newcommand{\texthide}[1]{}
\usepackage{xspace}
\usepackage{natbib}

\usepackage{subfigure}
\usepackage{savesym}
\savesymbol{tablenum}
\usepackage{siunitx}
\usepackage{amsmath}
\usepackage{datetime}
\usepackage{threeparttable}
\usepackage{tabularx}
\usepackage{longtable}
\usepackage{multirow}
\usepackage{graphicx}
\usepackage[super]{nth}

\restoresymbol{SIX}{tablenum}
\newdateformat{monthyeardate}{%
	\monthname[\THEMONTH] \THEYEAR}

\usepackage[encapsulated]{CJK}
\usepackage{ucs}
\usepackage[utf8x]{inputenc}
\newcommand{\cntext}[1]{\begin{CJK*}{UTF8}{gbsn}#1\end{CJK*}}
\newcommand{\jptext}[1]{\begin{CJK*}{UTF8}{min}#1\end{CJK*}}

\received{}
\revised{}
\accepted{}
\submitjournal{ApJ}
\shorttitle{}
\shortauthors{Wang et al.}

\begin{document}

\title{Architecture of planetary systems predicted from protoplanetary disks observed with ALMA I: mass of the possible planets embedded in the dust gap}

\correspondingauthor{Shijie Wang}
\email{shijie.wang@phys.s.u-tokyo.ac.jp}

\author[0000-0002-5635-2449]{Shijie Wang (\cntext{汪士杰})}
\affil{Department of Physics, The University of Tokyo, Tokyo 113-0033, Japan}

\author[0000-0001-7235-2417]{Kazuhiro D. Kanagawa (\jptext{金川和弘})}
\affiliation{Research Center for the Early Universe, School of Science, The University of Tokyo, Tokyo 113-0033, Japan}
\affiliation{College of Science, Ibaraki University, Mito 310-0056, Japan}

\author[0000-0002-4858-7598]{Yasushi Suto (\jptext{須藤靖})}
\affiliation{Department of Physics, The University of Tokyo,
  Tokyo 113-0033, Japan}
\affiliation{Research Center for the Early Universe, School of Science, The University of Tokyo, Tokyo 113-0033, Japan}

\begin{abstract}
	Recent ALMA observations have identified a variety of dust gaps in protoplanetary disks, which are commonly interpreted to be generated by unobserved planets.  Predicting mass of such embedded planets is of fundamental importance in comparing those disk architectures with the observed diversity of exoplanets. The prediction, however, depends on the assumption that whether the same gap structure exists in the dust component alone or in the gas component as well. We assume a planet can only open a gap in the gas component when its mass exceeds the pebble isolation mass by considering the core accretion scenario. We then propose two criteria to distinguish if a gap is opened in the dust disk alone or the gas gap as well when observation data on the gas profile is not available. We apply the criteria to 35 disk systems with a total of 55 gaps compiled from previous studies, and classify each gap into four different groups. The classification of the observed gaps allows us to predict the mass of embedded planets in a consistent manner with the pebble isolation mass.  We find that outer gaps are mostly dust alone, while inner gaps are more likely to be associated with a gas gap as well. The distribution of such embedded planets is very different from the architecture of the observed planetary systems, suggesting that the significant inward migration is required in their evolution.
\end{abstract}


\section{Introduction}
Protoplanetary disks (PPDs) around young stars are generally believed to be natural birthplaces for planets. In the standard core accretion scenario \citep[e.g.,][]{Mizuno1980,Hayashi1985}, dust grains inside the gas disk collide and coagulate, form rocky cores, and eventually grow to giant gaseous planets via runaway gas accretion. Alternatively, a fraction of giant planets may be formed {\it in-situ} by gravitational instability within a massive disk \citep[e.g.,][]{Cameron1978}. Planets in the PPDs are supposed to migrate in the disk via disk-planet interaction \citep[][]{Lin1979,Goldreich1980}, and their masses gradually increase due to the accretion of dust and gas components of the disk \citep[e.g.,][]{Tanigawa2002,DAngelo2003}. A large diversity of the architecture of the observed exoplanetary systems \citep[e.g.][]{Lissauer2011,Winn2014,Fabrycky2014} would be explained, at least partly, by such formation and evolution processes of planets in the disk phase.

A planet embedded in a disk gravitationally interacts with the surrounding materials and opens a surface density gap in the disk once it is sufficiently massive \citep[e.g.,][]{Lin1979,Goldreich1980,Crida2006}. The width and depth of the gas gap always increase with the planetary mass, and the gas pressure gradient at the vicinity of the planet is also modified as the gas gap is deepened. While a deep, fully-opened gas gap can only be opened by a massive planet, a planet that carves a shallow gas gap can already generate a pressure maximum outside of the gap. Since a negative pressure gradient is necessary for the dust grains to drift radially inwards, once a local pressure maximum is generated by the planet, the pebbles will be trapped and accumulated outside of the pressure maximum due to the inverse of the pressure gradient. \cite{Lambrechts2014} shows that a planet typically around $\sim \SI{20}{M_\oplus}$ is already sufficient to open a shallow gas gap with a gas pressure maximum, which is defined as the pebble-isolation mass. Conventionally, the dust gap was believed to open after the opening of the gas gap with the pressure maximum. However, recent studies suggest dust gaps can also be opened directly by a low mass planet smaller than the pebble-isolation mass, because dust is more susceptible to the gap opening torque than the gas \citep[e.g.][]{Dipierro2016, Rosotti2016, Dipierro2017}.
	
Recent observations by the Atacama Large Millimeter Array (ALMA) are able to spatially resolve the disk structure down to a few \si{au} scale, and have successfully identified various substructures in PPDs, including gaps and rings \citep[e.g.,][]{Partnership2015,Cieza2017,Huang2018a,Cieza2019,Kanagawa2021}, spiral arms \citep[e.g.,][]{Huang2018,Wolfer2021,Xie2021}, and inner cavities \citep[e.g.,][]{Dong2018a,Kudo2018,Hashimoto2021}. In particular, \cite{Long2018} reported 15 gaps in 12 disks and \cite{VanderMarel2019} identified 33 gaps or inner cavities in 16 sample disks. Such substructures are commonly interpreted as observational evidence for forming planets embedded in PPDs \citep[e.g.][]{Zhang2018,Bae2018,Lodato2019}. While alternative interpretations of gap and ring structures have been proposed, including gravitational instability \citep{Takahashi2016}, sintering \citep{Okuzumi2016}, snow lines \citep{Zhang2015} and MHD effects \citep{Hu2019}, the planetary interpretation is widely accepted as the most conventional model for the ring and gap substructure.

Following the planetary interpretation, the orbital semi-major axis and the mass of the hidden planet can be inferred from the location and width of the gap, respectively. More specifically, the planetary masses are predicted using the scaling relations between the gap width and planetary mass suggested by the numerical simulations \citep{Kanagawa2015,Kanagawa2016,Rosotti2016,Dong_Fung2017,Zhang2018}. In turn, the primordial architectures of planetary systems hosted by those disks can be reasonably constrained by the observed gap and ring structures. One may adopt them as empirically inferred initial conditions, simulate the evolution of those systems, which may be compared with various statistical properties of observed exoplanets \citep[e.g.,][]{Wang2020}. In principle, this provides an empirical methodology to link the disk data to the architecture of the observed planetary system, complementary to the conventional approach based on the purely theoretical initial conditions.

Since the gap opening mechanisms on gas and dust are physically different, the mass-width scaling relations in the respective regimes are likely to have different dependence on the planet and disk properties. Indeed, the scaling relations derived for massive planets that open both dust and gas gaps \citep{Kanagawa2016,Dong_Fung2017,Zhang2018} are found to have more explicit dependence on gas disk parameters than those derived for relatively low mass planets opening gaps only in the dust disk \citep{Dipierro2015,Rosotti2016,Dipierro2017}. The difference between those two models results in fairly different mass predictions and eventually leads to distinct fates for those disk systems, therefore posing a need to distinguish between the two gap opening scenarios. However, most of the observed gap/ring shapes can only be derived from the dust continuum image alone, so it is not clear if the gap is opened in the gas disk as well. Although a high angular resolution observation of gas line emission can resolve this degeneracy, it is technically difficult and feasible only for very nearby disks even with ALMA except for some particular nearby disks such as TW Hya and PDS 70 \citep{Nomura2021,ZCLong2018}.

The above difficulty motivates us to use mass criteria to differentiate the two scenarios and improve the planetary mass predictions. Theoretically, a planetary core formed by the collision or the fragmentation is supposed to grow via pebble accretion until its mass reaches the pebble-isolation mass $M_{\rm iso} (\sim \SI{10}{M_{\oplus}})$. \citep[e.g.,][]{Morbidelli2012,Lambrechts2014,Bitsch2018}. When the core mass is smaller than $M_{\rm iso}$, it forms no gap or very shallow gap in the gas disk. After the core mass exceeds $M_{\rm iso}$, the planetary core acquires a gas envelope due to the onset of runaway gas accretion \citep[e.g.,][]{Mizuno1980,Bodenheimer1986,Ikoma2000}. The resulting giant gaseous planet carves a deep gap in both gas and dust disks. According to the above theoretical understanding, the pebble isolation mass can be used as a boundary to break the degeneracy of two possible situations: the case that only the dust gap is opened, and the case that both the gas and dust gaps are opened.

In the present paper, we consider the two different predictions for the planetary mass from the dust gap properties, and discuss how we can distinguish between the two models, the dust only gap and both the gas and dust gaps. The rest of the paper is organized as follows. Section~\ref{sec:disk_gap_properties} describes disk systems that we consider in the present study, and summarizes the observed properties of observed gaps revealed by the dust continuum.  In section~\ref{sec:method}, we review the current understanding of the planet formation in the core-accretion scenario, the methods to deduce the planet mass from the shape of the gap, and the criteria to classify the gaps according to the two models. Section~\ref{sec:result} presents the result of gap classification and the corresponding planetary mass predicted from the classification. Section~\ref{sec:discussion} discusses the implications of our results on the planet formation and the caveats as well. Finally, section~\ref{sec:summary} is devoted to the summary and conclusion of this paper.

\section{Properties of disks and gaps of our sample}
\label{sec:disk_gap_properties}
Recent observations have discovered a number of substructures in PPDs. In particular, the continuum data of 20 nearby PPDs by The Disk Substructures at High Angular Resolution Project (DSHARP) \citep{Andrews2018} have revealed clear gap structures in most of the disks \citep{Huang2018a}. \cite{Long2018} surveyed 32 PPDs observed by ALMA in the Taurus star-forming region and identified gap and ring structures in 12 disks. Using ALMA archive data, \cite{VanderMarel2019} also identified gaps for 16 diverse disk systems. Our sample is assembled from previously published data. Since we are interested in gaps that are possibly induced by planets, we select those gaps that are likely to be of planetary origin following \cite{Zhang2018} and \cite{Lodato2019}. We also include the gaps in the HL Tau disk \citep{Partnership2015}. Our sample comprises of \num{35} disks with a total of \num{55} gaps.
\begin{table*}[!ht]
	\centering
	\caption{Properties of the disks and gaps}
	\resizebox{0.9\linewidth}{!}{
		\begin{threeparttable}
			\begin{tabular}{lllcccllllc}
				\hline\hline
				Disk& $ M_* $\tnote{a} & $ L_* $& $ T_{gas}(\SI{100}{au}) $& $ h/R (\SI{100}{au})$& \multirow{2}{*}{Gap index} & $ R_{\rm gap} $ & $ \Delta_{\rm gap} $ & \multirow{2}{*}{$ \Delta_{\rm gap}/R_{\rm gap} $} & \multirow{2}{*}{$ {h_{\rm gap}}/{R_{\rm gap}} $}&\multirow{2}{*}{Ref}\\
				&($  \si{M_\odot}  $)&($ \si{L_\odot} $)& (\si{K})&	&			    &	(\si{au})  & (\si{au}) & &	&\\
				\hline
				\textit{Single-gap disks}&&&&\\
				DM Tau          & $0.35^{+0.05}_{-0.05}$ & $0.2^{+0.03}_{-0.02}$   & 8.1  & 0.097 & 1 & 70.0  & 30.0 & 0.429 & 0.088 & 4     \\
				DN Tau          & $0.52$                 & $0.69$                  & 11.0 & 0.093 & 1 & 49.3  & 3.8  & 0.078 & 0.078 & 3     \\
				DS Tau          & $0.58$                 & $0.25$                  & 8.5  & 0.077 & 1 & 32.9  & 27.0 & 0.821 & 0.058 & 3     \\
				Elias 20        & $0.48^{+0.15}_{-0.07}$ & $2.24^{+1.31}_{-0.83}$  & 14.8 & 0.112 & 1 & 25.1  & 3.5  & 0.140 & 0.079 & 1,2   \\
				Elias 27        & $0.49^{+0.2}_{-0.11}$  & $0.91^{+0.64}_{-0.37}$  & 11.8 & 0.099 & 1 & 69.1  & 14.3 & 0.207 & 0.090 & 1,2   \\
				FT Tau          & $0.34$                 & $0.15$                  & 7.5  & 0.095 & 1 & 24.8  & 4.8  & 0.195 & 0.067 & 3     \\
				GW Lup          & $0.46^{+0.12}_{-0.15}$ & $0.33^{+0.19}_{-0.12}$  & 9.2  & 0.090 & 1 & 74.3  & 12.1 & 0.163 & 0.084 & 1,2   \\
				HD 135344B      & $1.5^{+0.1}_{-0.1}$    & $6.7^{+1.3}_{-2.9}$     & 19.4 & 0.072 & 1 & 73.0  & 6.0  & 0.082 & 0.067 & 4     \\
				HD 142666       & $1.58^{+0.15}_{-0.04}$ & $9.12^{+5.67}_{-3.5}$   & 21.0 & 0.073 & 1 & 16.0  & 3.5  & 0.219 & 0.046 & 1,2   \\
				HD 169142       & $1.75^{+0.05}_{-0.05}$ & $14^{+3.5}_{-5}$        & 23.4 & 0.074 & 1 & 42.0  & 20.0 & 0.476 & 0.059 & 4     \\
				HD 97048        & $2.2^{+0.1}_{-0.1}$    & $31.2^{+2.1}_{-6}$      & 28.6 & 0.072 & 1 & 125.0 & 90.0 & 0.720 & 0.077 & 4     \\
				IM Lup          & $0.89^{+0.21}_{-0.23}$ & $2.57^{+1.5}_{-0.95}$   & 15.3 & 0.083 & 1 & 117.4 & 15.8 & 0.135 & 0.087 & 1,2   \\
				IQ Tau          & $0.5$                  & $0.21$                  & 8.2  & 0.081 & 1 & 41.2  & 6.9  & 0.169 & 0.065 & 3     \\
				MWC 480         & $1.91$                 & $17.38$                 & 24.7 & 0.072 & 1 & 73.4  & 33.3 & 0.454 & 0.067 & 3     \\
				RU Lup          & $0.63^{+0.2}_{-0.14}$  & $1.45^{+0.85}_{-0.53}$  & 13.3 & 0.092 & 1 & 29.1  & 4.5  & 0.155 & 0.068 & 1,2   \\
				RXJ 1615.3-3255 & $0.6^{+0.1}_{-0.1}$    & $0.83^{+0.07}_{-0.07}$  & 11.5 & 0.088 & 1 & 92.5  & 35.0 & 0.378 & 0.087 & 4     \\
				RY Tau          & $2.04$                 & $12.3$                  & 22.6 & 0.067 & 1 & 43.4  & 4.9  & 0.112 & 0.054 & 3     \\
				SR 4            & $0.68^{+0.19}_{-0.19}$ & $1.17^{+0.69}_{-0.43}$  & 12.6 & 0.087 & 1 & 11.0  & 6.3  & 0.573 & 0.059 & 1,2   \\
				Sz 114          & $0.17^{+0.04}_{-0.03}$ & $0.2^{+0.12}_{-0.08}$   & 8.1  & 0.139 & 1 & 38.6  & 4.3  & 0.111 & 0.108 & 1,2   \\
				Sz 129          & $0.83^{+0.06}_{-0.24}$ & $0.44^{+0.26}_{-0.16}$  & 9.8  & 0.069 & 1 & 41.0  & 4.1  & 0.100 & 0.055 & 1,2   \\
				Sz 98           & $0.6^{+0.1}_{-0.1}$    & $1.1^{+0.2}_{-0.3}$     & 12.4 & 0.091 & 1 & 79.0  & 18.0 & 0.228 & 0.086 & 4     \\
				UZ Tau E        & $0.39$                 & $0.35$                  & 9.3  & 0.098 & 1 & 69.1  & 7.3  & 0.106 & 0.089 & 3     \\
				V1247 Ori       & $1.95^{+0.15}_{-0.15}$ & $20^{+2.7}_{-5.3}$      & 25.6 & 0.073 & 1 & 107.5 & 15.0 & 0.140 & 0.074 & 4     \\
				\hline
				\textit{Multi-gap disks}&&&&\\
				AA Tau          & $0.65^{+0.15}_{-0.15}$ & $0.56^{+0.2}_{-0.2}$    & 10.5 & 0.081 & 1 & 72.0  & 14.0 & 0.194 & 0.074 & 4     \\
				&                        &                         &      &       & 2 & 118.0 & 16.0 & 0.136 & 0.084 &       \\
				DoAr 25         & $0.95^{+0.09}_{-0.34}$ & $0.95^{+0.56}_{-0.35}$  & 11.9 & 0.071 & 1 & 98.0  & 15.5 & 0.158 & 0.071 & 1,2   \\
				&                        &                         &      &       & 2 & 125.0 & 10.0 & 0.080 & 0.075 &       \\
				Elias 24        & $0.85^{+0.05}_{-0.05}$ & $6.8^{+5.8}_{-3.2}$     & 19.5 & 0.096 & 1 & 55.0  & 20.0 & 0.364 & 0.083 & 4     \\
				&                        &                         &      &       & 2 & 94.0  & 12.0 & 0.128 & 0.095 &       \\
				GO Tau          & $0.36$                 & $0.21$                  & 8.2  & 0.096 & 1 & 58.9  & 22.9 & 0.389 & 0.084 & 3     \\
				&                        &                         &      &       & 2 & 87.0  & 16.1 & 0.185 & 0.093 &       \\
				HD 143006       & $1.78^{+0.22}_{-0.3}$  & $3.8^{+1.57}_{-1.11}$   & 16.9 & 0.062 & 1 & 22.0  & 21.7 & 0.986 & 0.042 & 1,2   \\
				&                        &                         &      &       & 2 & 51.0  & 12.8 & 0.251 & 0.052 &       \\
				AS 209          & $0.83^{+0.24}_{-0.23}$ & $1.41^{+0.83}_{-0.52}$  & 13.2 & 0.080 & 1 & 8.7   & 4.7  & 0.541 & 0.043 & 1,2,4 \\
				&                        &                         &      &       & 2 & 60.8  & 15.5 & 0.255 & 0.071 &       \\
				&                        &                         &      &       & 3 & 93.0  & 18.0 & 0.194 & 0.079 &       \\
				CI Tau          & $0.89$                 & $0.81$                  & 11.5 & 0.072 & 1 & 13.9  & 8.9  & 0.636 & 0.044 & 3     \\
				&                        &                         &      &       & 2 & 48.4  & 10.9 & 0.225 & 0.060 &       \\
				&                        &                         &      &       & 3 & 119.0 & 22.1 & 0.186 & 0.075 &       \\
				DL Tau          & $0.98$                 & $0.65$                  & 10.8 & 0.067 & 1 & 39.3  & 6.7  & 0.170 & 0.053 & 3     \\
				&                        &                         &      &       & 2 & 67.0  & 13.8 & 0.207 & 0.061 &       \\
				&                        &                         &      &       & 3 & 88.9  & 25.9 & 0.292 & 0.065 &       \\
				GY 91           & $0.19^{+0.01}_{-0.01}$ & $0.33^{+0.2}_{-0.2}$    & 9.2  & 0.140 & 1 & 41.0  & 14.0 & 0.341 & 0.112 & 4     \\
				&                        &                         &      &       & 2 & 69.0  & 10.0 & 0.145 & 0.127 &       \\
				&                        &                         &      &       & 3 & 107.0 & 14.0 & 0.131 & 0.142 &       \\
				HL Tau          & $1.0$                  & $9.25^{+5.5}_{-5.5}$    & 21.1 & 0.092 & 1 & 11.8  & 9.5  & 0.809 & 0.054 & Mix   \\
				&                        &                         &      &       & 2 & 32.3  & 7.5  & 0.233 & 0.070 &       \\
				&                        &                         &      &       & 3 & 82.0  & 24.0 & 0.293 & 0.088 &       \\
				V1094 Sco       & $0.8^{+0.1}_{-0.1}$    & $0.57^{+0.07}_{-0.13}$  & 10.5 & 0.073 & 1 & 60.0  & 10.0 & 0.167 & 0.064 & 4     \\
				&                        &                         &      &       & 2 & 102.5 & 35.0 & 0.341 & 0.073 &       \\
				&                        &                         &      &       & 3 & 174.0 & 20.0 & 0.115 & 0.084 &       \\
				HD 163296       & $2.04^{+0.25}_{-0.14}$ & $16.98^{+16.9}_{-8.47}$ & 24.5 & 0.070 & 1 & 10.0  & 3.2  & 0.320 & 0.039 & 1,2   \\
				&                        &                         &      &       & 2 & 48.0  & 20.2 & 0.421 & 0.058 &       \\
				&                        &                         &      &       & 3 & 86.4  & 16.2 & 0.188 & 0.067 &       \\
				&                        &                         &      &       & 4 & 145.0 & 13.4 & 0.092 & 0.077 &      \\
				\hline
			\end{tabular}
			\begin{tablenotes}[para,flushleft]
				\item[a] For HL Tau, we adopt the same stellar mass as \cite{Wang2020} adopted.\\
				\textbf{Reference.} (1) \cite{Andrews2018}; (2) \cite{Huang2018a}; (3) \cite{Long2018}; (4) \cite{VanderMarel2019}.
			\end{tablenotes}
		\end{threeparttable}
	}
	\label{tab:disk_gap_properties}
\end{table*}

Basic properties of stars, disks and gaps of our sample are summarized in Table~\ref{tab:disk_gap_properties}.  Stellar properties of disks listed in \cite{Zhang2018} and \cite{Lodato2019} are based on \cite{Andrews2018} and \cite{Long2018}, respectively. For disks sampled by \cite{VanderMarel2019}, we use the same stellar properties except for those overlapped with the DSHARP disks. For HL Tau system, we keep using the same parameters as we adopted in our previous study \cite{Wang2020}.

The stellar parameters can be used to estimate the disk temperature and the aspect ratio at a reference radius of \SI{100}{au} following the temperature scaling relation by \cite{Dong_Najita2018}. We assume that both dust and gas components of the disk have approximately the same temperature $T_{\rm dust} \approx T_{\rm gas}$. Then, the disk temperature at radius R is given by

\begin{equation}
	\label{eqn:dust_temp_fit} T(R) = \SI{12.1}{K}\left(\frac{L_*}{\rm L_\odot}\right)^{1/4} \left(\frac{R}{\SI{100}{au}}\right)^{-1/2}, 
\end{equation} 

where $L_*$ is the stellar luminosity.  Equation (\ref{eqn:dust_temp_fit}) is shown to be consistent with the results produced by previous radiative transfer models \citep[e.g.][]{Dong2015,Facchini2017}. We would like to note here that the $ T_{\rm dust} $ defined by equation (2) in the Appendix B of \cite{Dong_Najita2018} indeed corresponds to the dust temperature at $ R = 0.5 R_{\rm mm} $ instead of $ R = R_{\rm mm} $. Thus, our equation~(\ref{eqn:dust_temp_fit}) is identical to their model (R. Dong, personal communication).

For locally isothermal disks where the ideal gas equation of state holds, the gas aspect ratio at $ R $ is then

\begin{align} 
  \frac{h(R)}{R} &=
  \left(\frac{k_B T}{\mu m_p}\right)^{1/2} \Omega^{-1}_{K} R^{-1}\notag\\
  &=0.07\left(\frac{L_*}{L_\odot}\right)^{1/8}
  \left(\frac{M_*}{M_\odot}\right)^{-1/2}
  \left(\frac{R}{\SI{100}{au}}\right)^{1/4}, 
\end{align} 

where $\mu$ is the mean molecular weight (we adopt $\mu = 2.3$ in this paper), and $ m_p $ is the proton mass.

Table~\ref{tab:disk_gap_properties} also lists the properties of the gaps based on \cite{Huang2018a}, \cite{Long2018}, and \cite{VanderMarel2019}. Again for overlapped systems (e.g. AS 209 and HD 163296), we adopt DSHARP data that have a higher angular resolution. The third gap of the AS209 disk is adopted from \cite{VanderMarel2019}, which combines the fine gap structure found by DSHARP into a single gap centered at \SI{93}{au}. For HL Tau, we keep using the same gap properties as we adopted in \cite{Wang2020}.

Among the 35 disks, 23 disks are single-gap systems, and the rest are multi-gap systems. The width of the gap $\Delta_{\rm gap}$ is defined to be the difference between the outer and inner edges for systems adopted from \cite{VanderMarel2019}, and the gap location is set to be the midpoint of the outer and inner edges. We neglect all the inner cavities (i.e. gaps start at \SI{0}{au}), because the interpretation of such a structure remains unclear, making it difficult to deduce the mass or number of the potential planet(s) inside the cavity. We also calculate $ \Delta_{\rm gap}/R_{\rm gap} $ and the gas aspect ratio $ h_{\rm gap}/R_{\rm gap} $. Most of the gaps have widths greater than the scale heights, with the third gap of GY 91 as the only exception.

\section{Methods}
\label{sec:method}

\begin{figure*}[thb]
	\centering
	\includegraphics[width=0.8\linewidth]{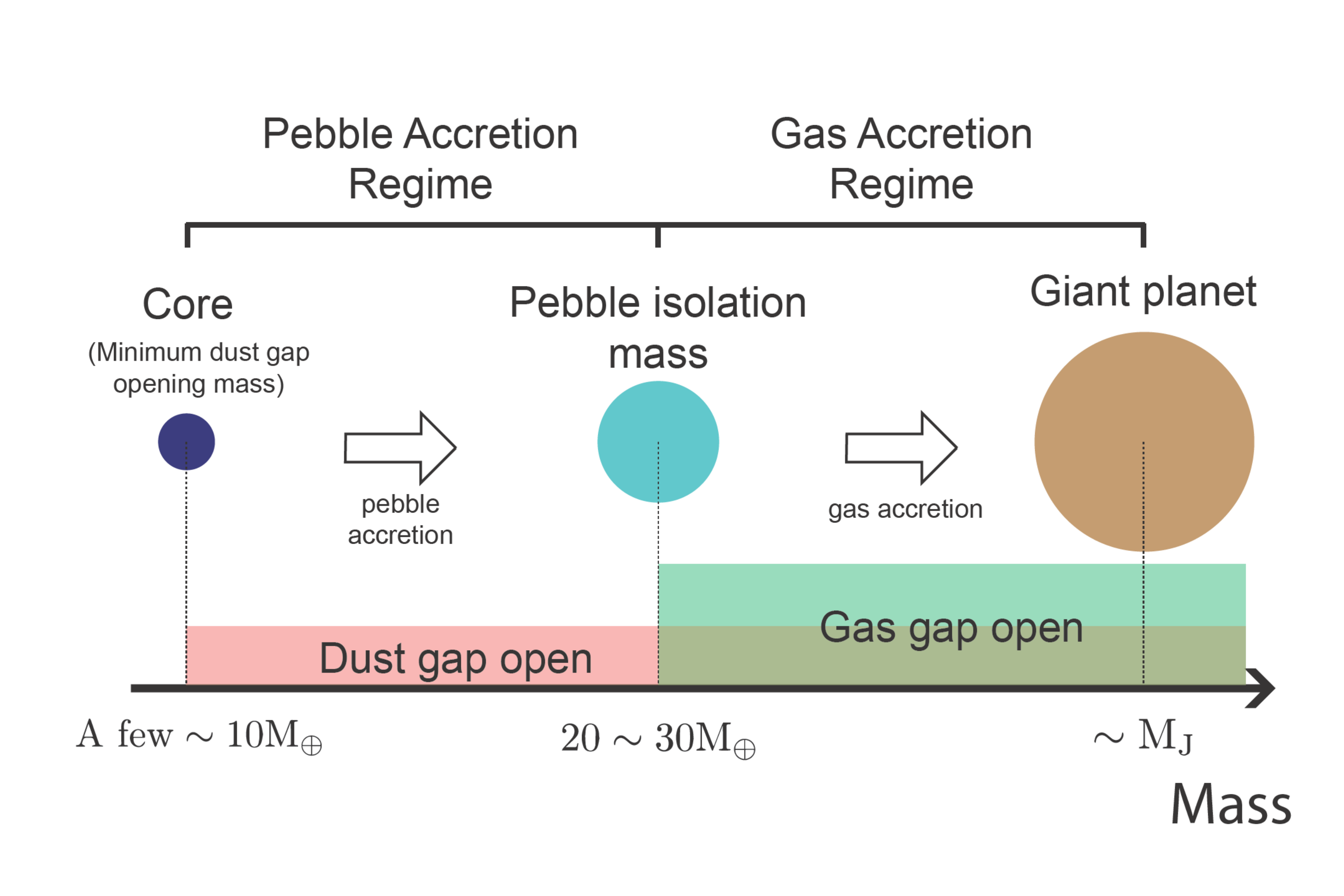}   
	\caption{Schematic mass growth history of planets we considered in this study. A solid core is formed first and then grows via pebble accretion. The dust gap is opened at the minimum dust gap opening mass. After reaching the pebble isolation mass $M_{\rm iso}(R)$ at the location of the planet $R$, the pebble accretion process terminates and the gas accretion begins.} \label{fig:overview_massgrowth}
\end{figure*}

\subsection{Mass accretion history of planets and pebble isolation mass}
\label{sec:planet_formation_picture}

In the standard core-accretion scenario, a giant planet forms out of a small solid core, and grows via two-stage accretion processes as illustrated in Figure \ref{fig:overview_massgrowth}. At the first stage when its core mass is small, a planet mainly accretes solid materials from the surrounding disk. While it was believed that the accreting material at this stage is mainly \si{km}-size planetesimals \citep[e.g.][]{Safronov1969}, recent studies \citep[e.g.][]{Ormel2010,Ormel2017} suggested that the aerodynamically small particles with Stokes number between \num{e-3} and \num{1}, so-called pebbles, may instead be the major source of accretion. In the following, we will use the terms "dust grains" and "pebbles" interchangeably. 

As pebble accretion continues, the planet becomes more massive and starts to gravitationally interact with its surrounding disk with increasing strength. Several studies have suggested that the dust disk is more susceptible to gap opening torque than gas disk \citep[e.g.][]{Paardekooper2004,Muto2009,Rosotti2016}. Before the planet is massive enough to generate a pressure maximum, it is possible to open a gap only in the dust disk through tidal torque and gas drag \citep{Dipierro2016,Dipierro2017}.
For instance, \cite{Rosotti2016} showed that even a planet with $\SI{15}{M_\oplus}$ can open the dust gap without creating the pressure maximum. This result implies that dust gaps carved by planets above the minimum dust gap opening mass can be possibly observed by ALMA that is sensitive to millimeter-size dust particles. Once the dust gap is opened, the planet loses access to the fraction of pebbles sensitive to the ALMA band. Nevertheless, until the planet reaches the pebble isolation mass $M_{\rm iso}$ and generates a gas pressure maximum, some pebbles may still cross the planetary orbit depending on their Stokes number.

The pebble accretion is a self-terminating process that ends when the planetary mass reaches a certain mass threshold, namely the pebble isolation mass \citep[e.g.][]{Lambrechts2014}. At the pebble isolation mass, a planet is just massive enough to carve a shallow gas gap that generates a pressure maximum at the outer edge of the gap. Such a pressure maximum can trap \textit{all} the pebbles with Stokes number $ > 0.005 $ \citep{Bitsch2018}, a range that effectively covers those pebbles that are most favourable to pebble accretion. The planet has no access to the inflow of the pebbles afterwards and the pebble accretion process is stopped.
Since both the width and depth of the gas gap increase with the planetary mass, it is reasonable to expect that a relatively deep gas gap is formed once the planetary mass exceeds the pebble-isolation mass, and the dust gap is also strongly shaped by the gas gap due to the pressure maximum. Before reaching the pebble-isolation, it is reasonable to assume that a visible gap should be formed only in the dust disk because the gas gap is too shallow.

Three-dimensional hydrodynamical simulations by \citep{Bitsch2018} indicate the pebble isolation mass $M_{\rm iso}(R)$ at the location of $R$ from the central star, is written in terms of the stellar mass $M_*$, the aspect ratio of the disk $h/R$, the pressure gradient $\eta = \partial\ln P/\partial\ln R$, and the viscosity of the gas $\alpha$ as
\begin{align}
	\label{eqn:peb_iso_mass}
	M_{\rm iso}(R) &= 25 M_{\oplus}\left(\frac{M_*}{M_\odot}\right)
	\left(\frac{h/R}{0.05}\right)^3f(\alpha,\eta), \\
	\label{eq:f-alpha-eta}
	f(\alpha,\eta) &=
	\left[0.34\left(\frac{-3.0}{\log\alpha}\right)^4+0.66\right]
	\left(1-\dfrac{\eta + 2.5}{6}\right).
\end{align}

Throughout the present paper, we consider locally isothermal disks with a temperature profile of $ T \propto R^{-1/2} $ and a surface density profile of $ \Sigma \propto R^{-1} $.  We simply adopt the mid-plane pressure $ P_{\rm mid} = \rho_{\rm mid} c_s^2 = (\Sigma/h) c_s^2$. Then we have $\eta = -2.75 $ and the second bracket in equation (\ref{eq:f-alpha-eta}) is $25/24$. It is worth to note that $M_{\rm iso}$ obtained from 2D simulations is scaled down by a factor of \num{1.5} due to the different treatments of the surface density profile \citep[][Appendix B]{Bitsch2018}. In this study, we nevertheless adopt $M_{\rm iso, 3D}$ as the pebble isolation mass as we think it is more reliable while applying to real systems.

After reaching the pebble isolation mass, the gas envelop around the planet starts cooling, which triggers the run-away gas accretion that marks the beginning of the second stage. At this stage, the planet grows quickly via gas accretion and becomes a gas giant. As the planet continues to grow, it carves a deeper gap in the disk, and the gas accretion rate is slowed down. The accretion eventually stops when the PPD is dispersed due to both stellar accretion and photoevaporation.

\subsection{Planetary mass estimation} 
\label{subsec:plt_mass_estimation}

The planetary formation picture illustrated in Figure \ref{fig:overview_massgrowth} states that a planet opens a gap when it is massive enough. Since the properties of the gap is determined by the planet-disk interaction, one may infer the mass of the embedded planet from the parameters characterizing the gap morphology.  Indeed, \cite{Kanagawa2016} performed a series of hydrodynamical simulations in the gas-gap open case, and found the following empirical relation for the mass of the embedded planet:
\begin{align}
	\label{eqn:pltmass_Kanagawa}
	M_{\rm p, gas}(R_{\rm gap}) = 175M_\oplus \left(\frac{M_*}{M_\odot}\right)
	\left(\frac{\Delta_{\rm gap}/R_{\rm gap}}{0.5}\right)^2
	\left(\frac{h_{\rm gap}/R_{\rm gap}}{0.05}\right)^{3/2}
	\left(\frac{\alpha}{\num{e-3}}\right)^{1/2},
\end{align}
where $\Delta_{\rm gap}$, $h_{\rm gap}$ and $R_{\rm gap}$ are the width, scale-height and location of the gap. The empirical formula, equation (\ref{eqn:pltmass_Kanagawa}), has been basically confirmed by \cite{Dong_Fung2017}.

It is important to note that $ \Delta_{\rm gap} $ in equation~(\ref{eqn:pltmass_Kanagawa}) is defined for the gas profile in the original study. Therefore, while applying this mass prediction formula to those gaps observed in dust continuum, we need to assume that the gas gap has already opened, and its width roughly matches that of the dust gap due to good gas-dust coupling. 

\cite{Zhang2018} carried out gas--dust hydrodynamic simulations and investigated the scaling relation between the planet mass and the width of the gap like equation~(\ref{eqn:pltmass_Kanagawa}).
Overall their fitting is consistent to Equation~(\ref{eqn:pltmass_Kanagawa}) \footnote{The formulation of \cite{Zhang2018} is different from equation~\ref{eqn:pltmass_Kanagawa}, but this different is mainly due to their own measurement of the gap width.}, though it is slightly deviated due to the decoupling between gas and dust when $St\gtrsim 10^{-2}$.
Due to this deviation, Equation~(\ref{eqn:pltmass_Kanagawa}) may overestimate the planet mass.
To take this difference into account, we include an uncertainly factor for the estimated mass as will be discussed in section~\ref{sec:gap_classification_improved_mass}.
Also as will be seen in section~\ref{subsec:plt_mass_estimation}, the estimated mass given by equation~(\ref{eqn:pltmass_Kanagawa}) agrees with that estimated by \cite{Zhang2018} within this level of uncertainly.

Before reaching the pebble-isolation mass, no significant gas gap is expected to form and only a dust gap is visible as described in section~\ref{sec:method}. 
Equation~(\ref{eqn:pltmass_Kanagawa}) is not relevant in order to estimate the mass of the embedded planet in such a dust-only gap regime. For instance, Figure 12 of \cite{Zhang2018} exhibits a large scatter of the range of small planetary mass (or small $ K^{\prime} $), which may possibly indicate the transition of the different scaling between the dust-only gap regime and gas gap regime that we assumed in this paper. In the dust-only gap regime, the width of the dust gap would be better described by the Hill radius scaling, instead of equation~(\ref{eqn:pltmass_Kanagawa}).

The Hill radius model suggests that the gap width $\Delta_{\rm gap}$ is proportional to the the planetary Hill radius $R_{\rm H}$ at the gap location $R_{\rm gap}$.  This relation can be rewritten for the planetary mass as
\begin{equation}
	\label{eqn:pltmass_Hill}
	M_{\rm p, dust}(R_{\rm gap})= 6.01M_\oplus
	\left(\frac{C}{5.5}\right)^{-3}\left(\frac{M_*}{M_\odot}\right)
	\left(\frac{\Delta_{\rm gap}/R_{\rm gap}}{0.1}\right)^{3} ,
\end{equation}
where $C$ is the proportional constant and previous studies indicate that $4.5<C<7$ \citep{Rosotti2016}. Following \cite{Lodato2019}, we set $C= 5.5$ in this study. 

While equations~(\ref{eqn:pltmass_Kanagawa}) and (\ref{eqn:pltmass_Hill}) should connect to each other, the connection might not be smooth.
Indeed, \cite{Rosotti2016} investigated both the dust-only gap case and relatively deep gas gap case and found the gap width can be nicely scaled by a single Hill radius (see Figure~16 of \cite{Rosotti2016}).
Although their scaling result does not explicitly depend on any gas parameters, it implies that equation~(\ref{eqn:pltmass_Hill}) may be valid even in the gas-gap opening regime for some specific setup.
Investigation of the connection between the two scaling relations is beyond the scope of this paper and will be reserved for the future work.

The above argument indicates that equation~(\ref{eqn:pltmass_Kanagawa}) is applicable only for a gaseous planet embedded in a disk with both dust and gas gaps, while equation~(\ref{eqn:pltmass_Hill}) may be more relevant for a low mass planet with a dust gap alone. Unfortunately, the high-resolution disk substructure is available only for the dust component due to the difficulty of resolving the line emission from gas. Therefore, it is not easy to observationally distinguish between the two different predictions for the mass of embedded planets.

According to the planetary formation picture described in section~\ref{sec:planet_formation_picture}, however, the pebble isolation mass $M_{\rm iso}$ may be used as a reasonable threshold of the planetary mass that is consistent with the presence or absence of the unobservable gas gap; if the predicted planetary mass is greater than $M_{\rm iso}$, the gas gap is likely to be opened as well.

Equation~(\ref{eqn:pltmass_Kanagawa}) implies that $M_{\rm p, gas}>M_{\rm iso}(R_{\rm gap})$ is translated into the lower limit of $(\Delta_{\rm gap}/R_{\rm gap})$:
\begin{equation}
\label{eqn:crit_gas_width}
	\left(\frac{\Delta_{\rm gap}}{R_{\rm gap}}\right)_{\rm gas} > 
	0.189
		\left[\frac{M_{\rm iso}(R_{\rm gap})}{25M_\oplus}\right]^{1/2} 
		\left(\frac{M_*}{M_\odot}\right)^{-1/2}
		\left(\frac{h_{\rm gap}/R_{\rm gap}}{0.05}\right)^{-3/4}
		\left(\frac{\alpha}{\num{e-3}}\right)^{-1/4}.
\end{equation}
Thus, if $(\Delta_{\rm gap}/R_{\rm gap})$ of the observed {\it dust} gap satisfies the above inequality (\ref{eqn:crit_gas_width}), the gap is likely expected to be associated with the {\it gas} gap of the similar value of $(\Delta_{\rm gap}/R_{\rm gap})_{\rm gas}$. Thus equation~(\ref{eqn:pltmass_Kanagawa}) provides a reasonably good mass estimate of the embedded planet.

Similarly, equation~(\ref{eqn:pltmass_Hill}) implies that $M_{\rm p, dust}<M_{\rm iso}(R_{\rm gap})$ is rewritten as
\begin{equation}
	\label{eqn:crit_dust_width}
	\left(\frac{\Delta_{\rm gap}}{R_{\rm gap}}\right)_{\rm dust} <
	0.161
	\left[\frac{M_{\rm iso}(R_{\rm gap})}{25M_\oplus}\right]^{1/3}
	\left(\frac{M_*}{M_\odot}\right)^{-1/3}.
\end{equation}
If a dust gap satisfies the inequality (\ref{eqn:crit_dust_width}), it is likely to be a dust-only gap, and equation~(\ref{eqn:pltmass_Hill}) may be preferred to predict the mass of the embedded planet.

The above argument offers a possibility to distinguish between the presence and absence of the gas gap with the similar profile of the observed dust gap. In reality, however, the mass predictions suffer from significant uncertainties for multiple reasons. The major uncertainty comes from the gap width $\Delta_{\rm gap}$. \cite{Rosotti2016} found that the gap profile in their simulations changes in a time-dependent fashion, and the observed gap width would vary accordingly, which leads to a factor of \num{2}-\num{3} uncertainty for the planetary mass prediction. Also equation~(\ref{eqn:pltmass_Kanagawa}) derived by \cite{Kanagawa2016} depends on the coupling strength between dust and gas components in the disk, which would affect its overall normalization.  Furthermore, the definitions of the inner and outer boundaries of the gap are not unique, and vary among different papers.  Finally, the host stars in PPDs are generally young, and their luminosities and masses cannot be well constrained from the stellar evolution model. Therefore, we have to keep in mind such uncertainties in interpreting inequalities (\ref{eqn:crit_gas_width}) and~(\ref{eqn:crit_dust_width}).

\subsection{Gap classification for improved mass prediction}
\label{sec:gap_classification_improved_mass}
Due to various uncertainties of the observed dust profiles, we introduce fudge factors in the planetary mass predictions. Specifically, we impose the following criteria (hereafter ``C1" and ``C2") for the dust only gap and for the gas gap, respectively:
\begin{align}
&\textit{C1 }\text{(dust gap)}:\quad \frac{M_{\rm p, dust}}{M_{\rm iso} } < F_1
		\label{eqn:crit1}\\
&\textit{C2 } \text{(gas gap)}:\quad \frac{M_{\rm iso}}{M_{\rm p, gas}} <  F_2
		\label{eqn:crit2},
\end{align}
where $F_1$ and $F_2$ are the fudge factors. Substituting equations~(\ref{eqn:peb_iso_mass}), (\ref{eqn:pltmass_Kanagawa}) and (\ref{eqn:pltmass_Hill}), inequalities (\ref{eqn:crit1}) and (\ref{eqn:crit2}) are rewritten as the relations between ${\Delta_{\rm gap}}/{R_{\rm gap}}$ and ${h_{\rm gap}}/{R_{\rm gap}}$:
\begin{align}
	\textit{C1}:\quad \log\left(\frac{\Delta_{\rm gap}}{R_{\rm gap}}\right)
	&< \log\left(\frac{h_{\rm gap}}{R_{\rm gap}}\right)
	+ \frac{1}{3}\log f(\alpha) + \frac{1}{3}\log F_1 + 0.507 \\	
	\textit{C2}:\quad \log\left(\frac{\Delta_{\rm gap}}{R_{\rm gap}}\right)
	& > 0.75\log\left(\frac{h_{\rm gap}}{R_{\rm gap}}\right)
	+\frac{1}{2}\left[\log f(\alpha)-0.5\log\alpha\right]
	-\frac{1}{2}\log F_2 - 0.498,
\end{align}

\begin{figure}
	\centering
	\includegraphics[width=0.8\linewidth]{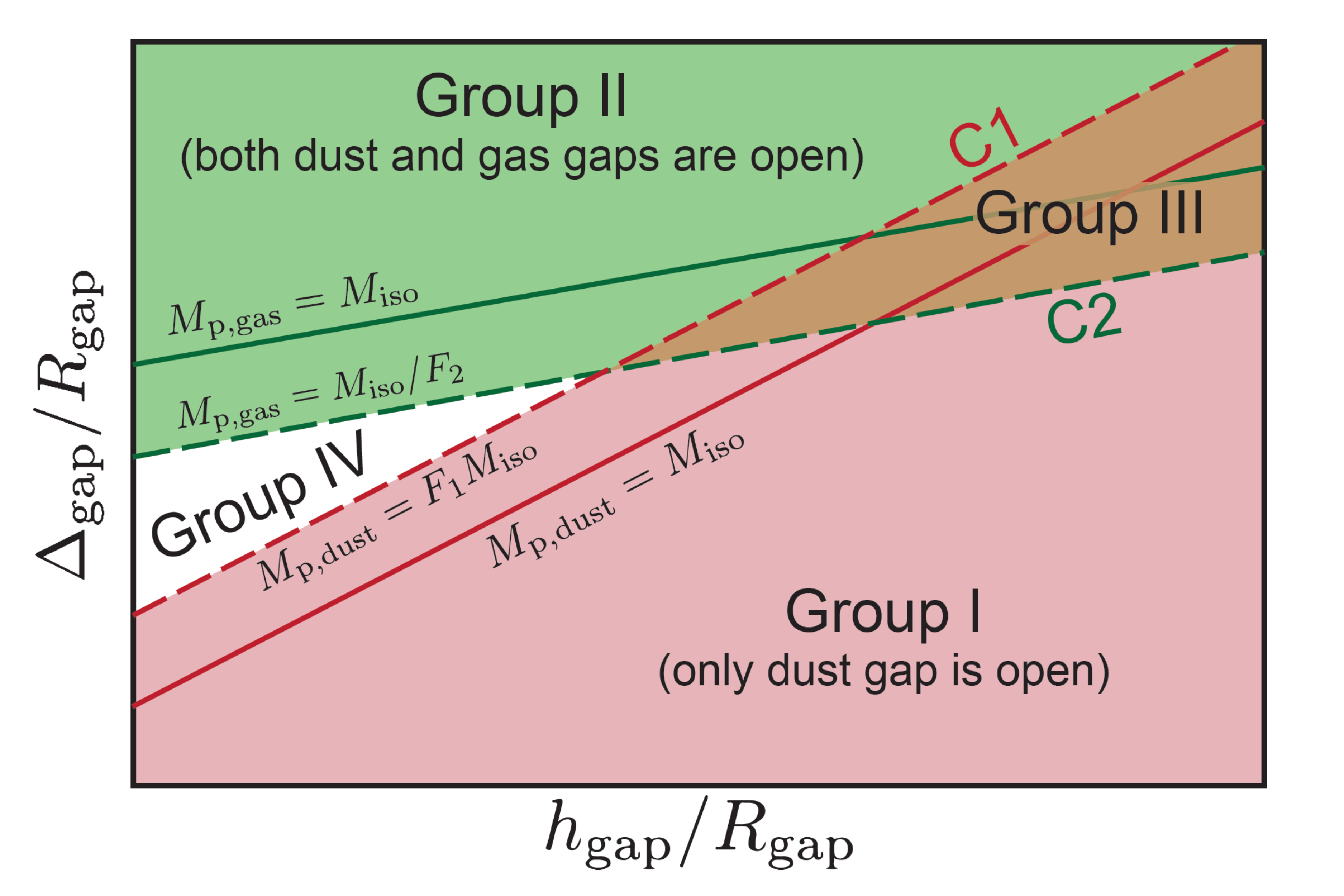}
	\caption{Schematic illustration of the gap classification}
	\label{fig:gpillustration}
\end{figure}

According to the two criteria C1 and C2, the observed gaps are classified into four groups on ${h_{\rm gap}}/{R_{\rm gap}}$ -- ${\Delta_{\rm gap}}/{R_{\rm gap}}$ plane, as schematically illustrated in Figure~\ref{fig:gpillustration} and summarized in Table \ref{tab:gap_classification}. While the relevant choice of the fudge factors is not clear, we adopt $F_1 = F_2 = 2$ below, just for simplicity.
 
 \begin{table*}
 	\centering
 	\caption{Criteria to classify the gaps into four groups.}
 	\begin{tabular}{lccccccc}
 		\hline
 		& \multirow{2}{*}{C1} & \multirow{2}{*}{C2} & \multirow{2}{*}{Interpretation} & \multirow{2}{*}{Mass adopted} & \multicolumn{3}{c}{Number of gaps} \\
 		&&&&&$ \alpha = \num{e-2} $& $ \alpha = \num{e-3} $& $ \alpha = \num{e-4} $\\
 		\hline
 		Group I & Yes & No & Dust only gaps & $ M_{\rm p,dust} $ &18&21&36\\
 		Group II & No & Yes & Dust/Gas gaps & $ M_{\rm p,gas} $ &11&19&19\\
 		Group III & Yes & Yes & Indistinguishable & $ M_{\rm p,dust} $, $ M_{\rm p,gas} $ &26&15&0\\
 		Group IV & No & No & Non-planetary? & - &0&0&0\\
 		\hline
 	\end{tabular}
 	\label{tab:gap_classification}
 \end{table*}
 
On the basis of the classification, we can interpret the nature of the gap. Group I implies that the observed dust gap is not associated with a gas gap, and equation~(\ref{eqn:pltmass_Hill}) is preferred for the planetary mass estimate. Group II, on the other hand, implies that the disk has a gas gap as well, and equation~(\ref{eqn:pltmass_Kanagawa}) is a reasonable estimate for the embedded gas planet. Gaps in group III are consistent with both interpretations and we cannot distinguish between the two possibilities $M_{\rm p, dust}$ and $M_{\rm p, gas}$, unless the presence or absence of the gas gap is clarified observationally.

Finally, gaps in group IV are inconsistent with either interpretation. Such gaps may be of non-planetary origin, or it may simply point to a possible inaccuracy or limitation of the mass prediction models that we adopted here. In this context, it is interesting to note that there is no gap classified as group IV in our sample. More details of the interpretation of the groups can be found in appendix \ref{sec:appendix}, where we show how the gap width varies with planetary mass at three different locations of a disk.

\section{Results}
\label{sec:result}
\subsection{Classification of the gaps}

\begin{figure*}
	\centering
	\gridline{\fig{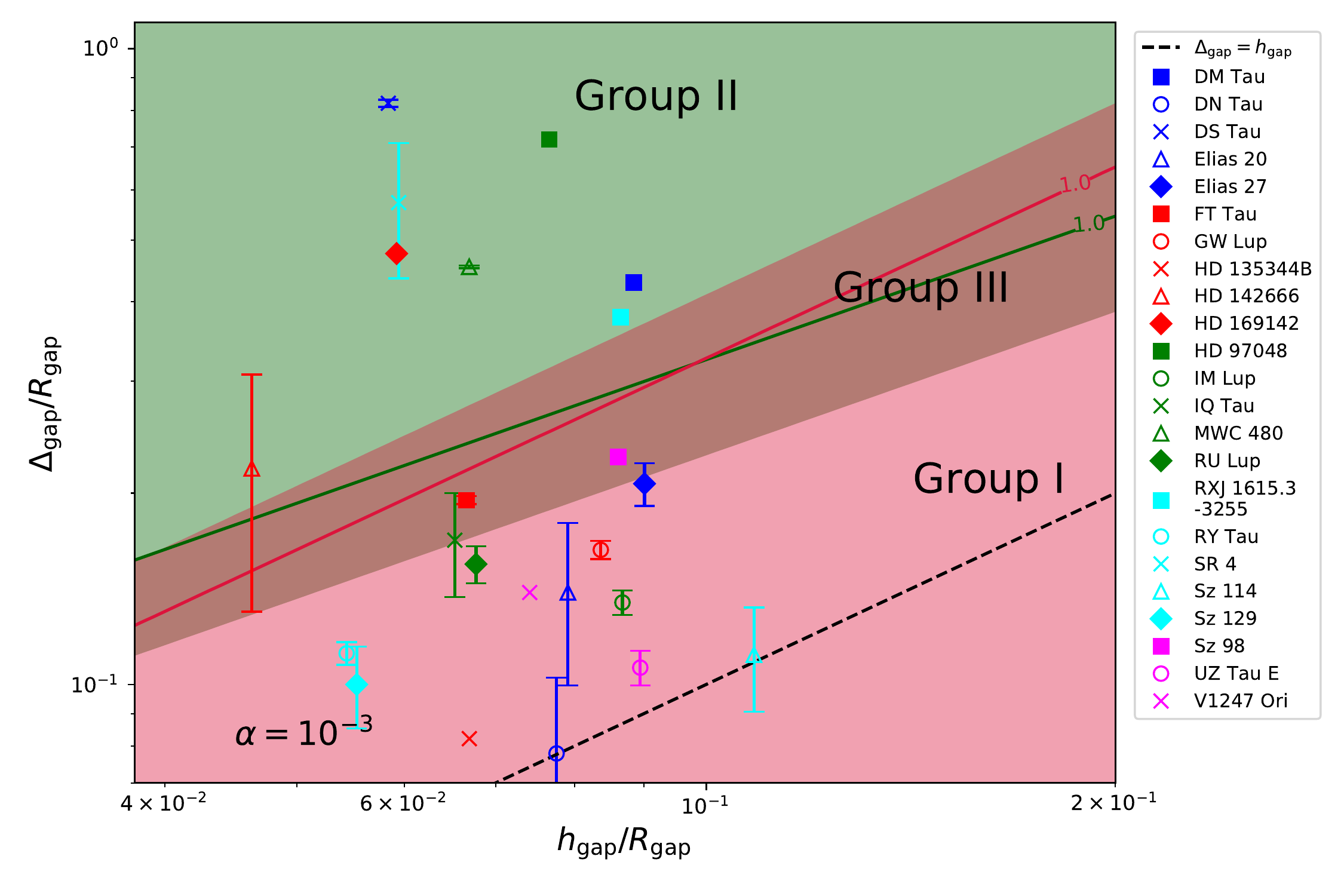}{0.8\linewidth}{(a) Single-gap disk systems}}
	\gridline{\fig{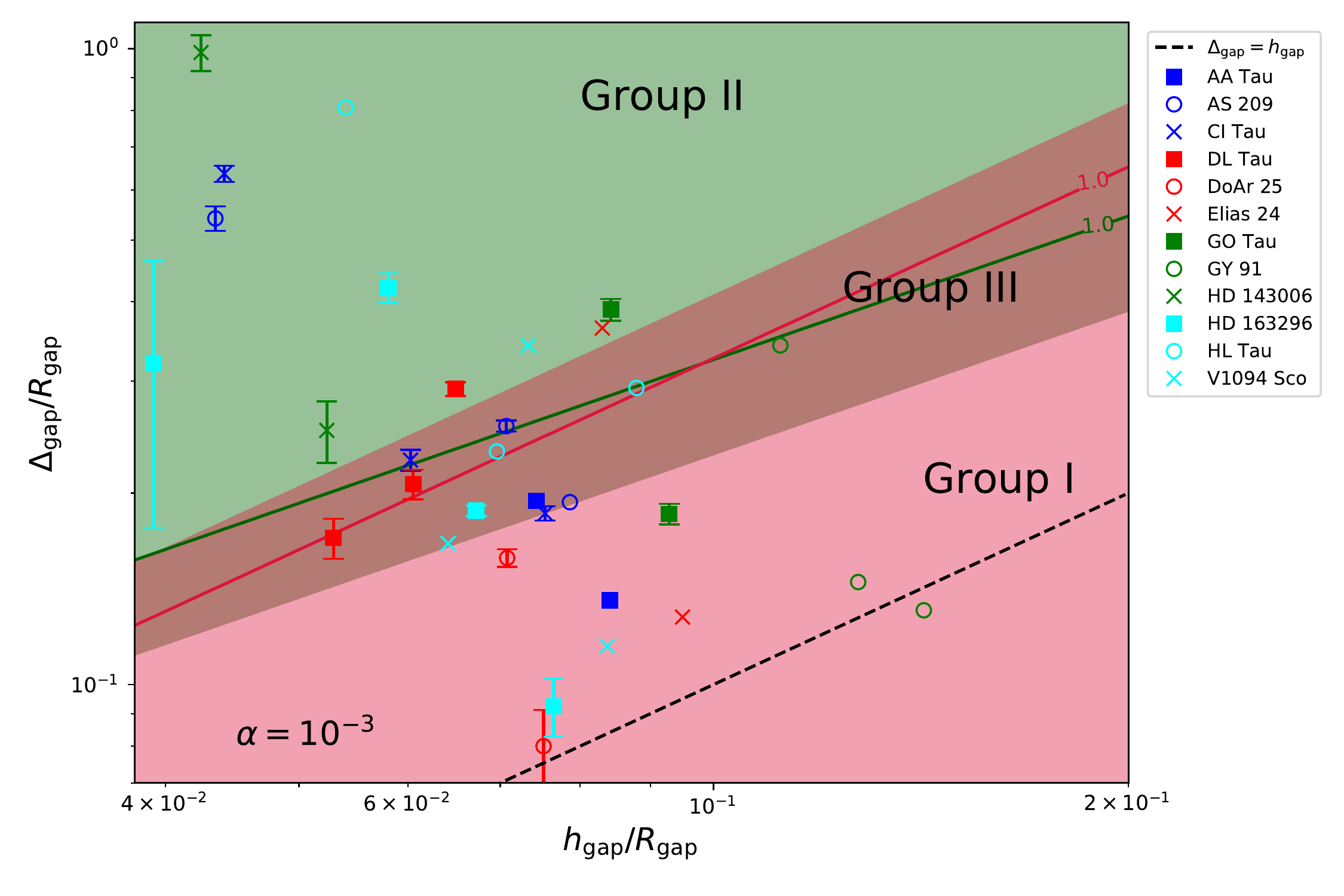}{0.8\linewidth}{(b) Multi-gap disk systems}}
	\caption{Classification of the gaps at $ \alpha = \num{e-3}
          $. Figure above shows the gaps from single-gap systems, and
          figure below show those from multi-gap systems. Group I, II
          and III correspond to red, green and brown regions. There is
          no group IV region at the given range of the data. The red
          line represents $ M_{\rm p, dust}/M_{\rm iso} = 1.0 $. The
          green line represents $ M_{\rm p, gas}/M_{\rm iso} = 1.0
          $. The black dashed line represents $\Delta_{\rm gap} =
          h_{\rm gap}$. The brown region is bounded by criteria C1
          (equation \ref{eqn:crit1}) and C2 (equation
          \ref{eqn:crit2}).}
	\label{fig:gap_classification_1e-3}
\end{figure*}	

According to C1 and C2, we classify the gaps into groups I to IV as shown in Figure~\ref{fig:gap_classification_1e-3}. The red and green regions correspond to group I (dust-gap only) and group II (gas-gap opened) gaps respectively, and the overlapped region (light-brown) bounded by equations~(\ref{eqn:crit1}) and (\ref{eqn:crit2}) is group III in which both interpretations are possible. We adopt the fiducial value of the viscosity, $\alpha=10^{-3}$.  The relations of $ M_{\rm p, dust} = M_{\rm iso} $ and $ M_{\rm p, gas} = M_{\rm iso} $ are plotted in red and green lines, respectively. The quoted error-bars of each symbol along the $y$-axis is estimated from the $1\sigma$ uncertainties of $ R_{\rm gap} $ and $ \Delta_{\rm gap} $ listed in previous studies. If they are not available, we do not plot the error-bar.

The result shows that out of 55 gaps, there are 21 gaps in group I, 19 gaps in group II and 15 gaps in group III, i.e., more than two-thirds ($ 73 \% $) of the gaps are classified as either dust-only gaps or gas gaps according to our criteria. The majority ($ 52 \% $) of the gaps in single-gap disks are classified as dust-only gaps, which are also the dominant group in the multi-gap disks. It should be noted that when the predicted planetary mass is close to $ M_{\rm iso} $, the gap classification becomes sensitive to the adopted model of temperature and aspect ratio. Only one gap is below the dashed line $ \Delta_{\rm gap} = h_{\rm gap} $, which is the minimum dust gap width for the Stokes number $ St = 1 $. If this gap is opened by a planet, the coupling between the dust and gas should be weak with $ St > 1 $ \citep{Dipierro2017,Lodato2019}. While we give mass predictions to this gap as well, it may be of non-planetary origin.

The distribution of the gaps from both single-gap disks and multi-gap disks shows that gaps with small aspect ratios (and thus at the inner region) are more likely to be gaps that are opened in both gas and dust disks. In the outer region where the aspect ratio is larger, the gaps are more likely to be dust alone. This trend is more clearly seen in the case of multi-gap disks (bottom panel of Figure \ref{fig:gap_classification_1e-3}), which may be attributed to presence of the planet-planet interaction in a multi-planetary system. We will discuss this difference further in section \ref{sec:implications}.

\subsection{Planetary mass predictions}

\begin{figure}
	\centering
	\includegraphics[width=\linewidth]{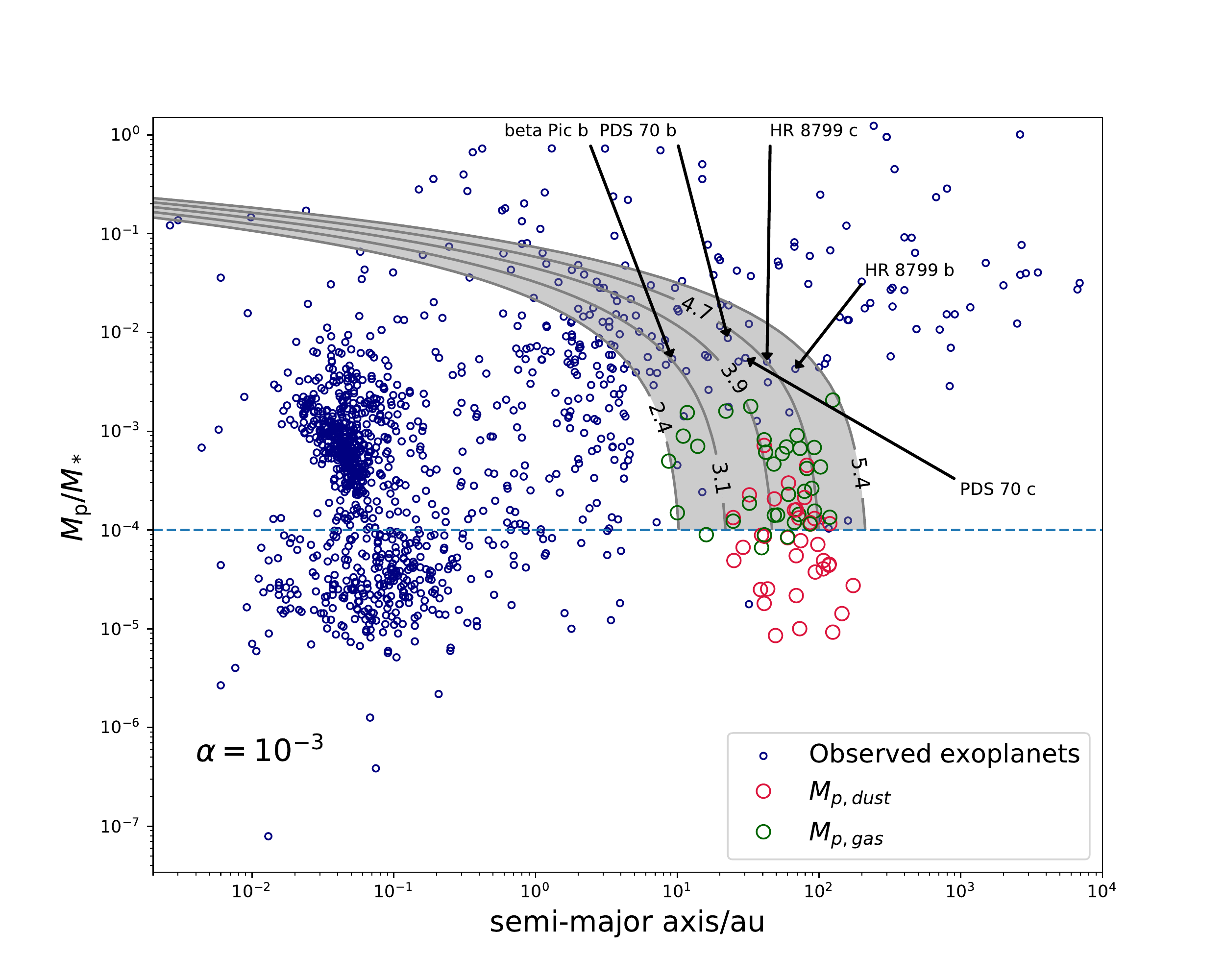}
	\caption{Compare the predicted masses ($ \alpha = \num{e-3} $) with the observed population of exoplanets whose semi-major axis and mass are both confirmed. The red dots are $ M_{\rm p,dust} $ when only dust gap is opened, while green dots are $ M_{\rm p,gas} $ when both gas/dust gaps have been opened. If a gap is in group III (indistinguishable), both predictions are plotted. The grey lines are the evolution tracks introduced by \cite{Tanaka2019}, and the number labels are constants $\mathcal{C} $ in equation (\ref{eqn:evo_track}). Since the evolution model is only applicable to gas accretion planets, we set the lower limit of the evolution tracks to be \num{e-4}.}
	\label{fig:comp_obs_mvsr}
\end{figure}

\begin{figure*}
	\centering
	\includegraphics[width=0.9\linewidth]{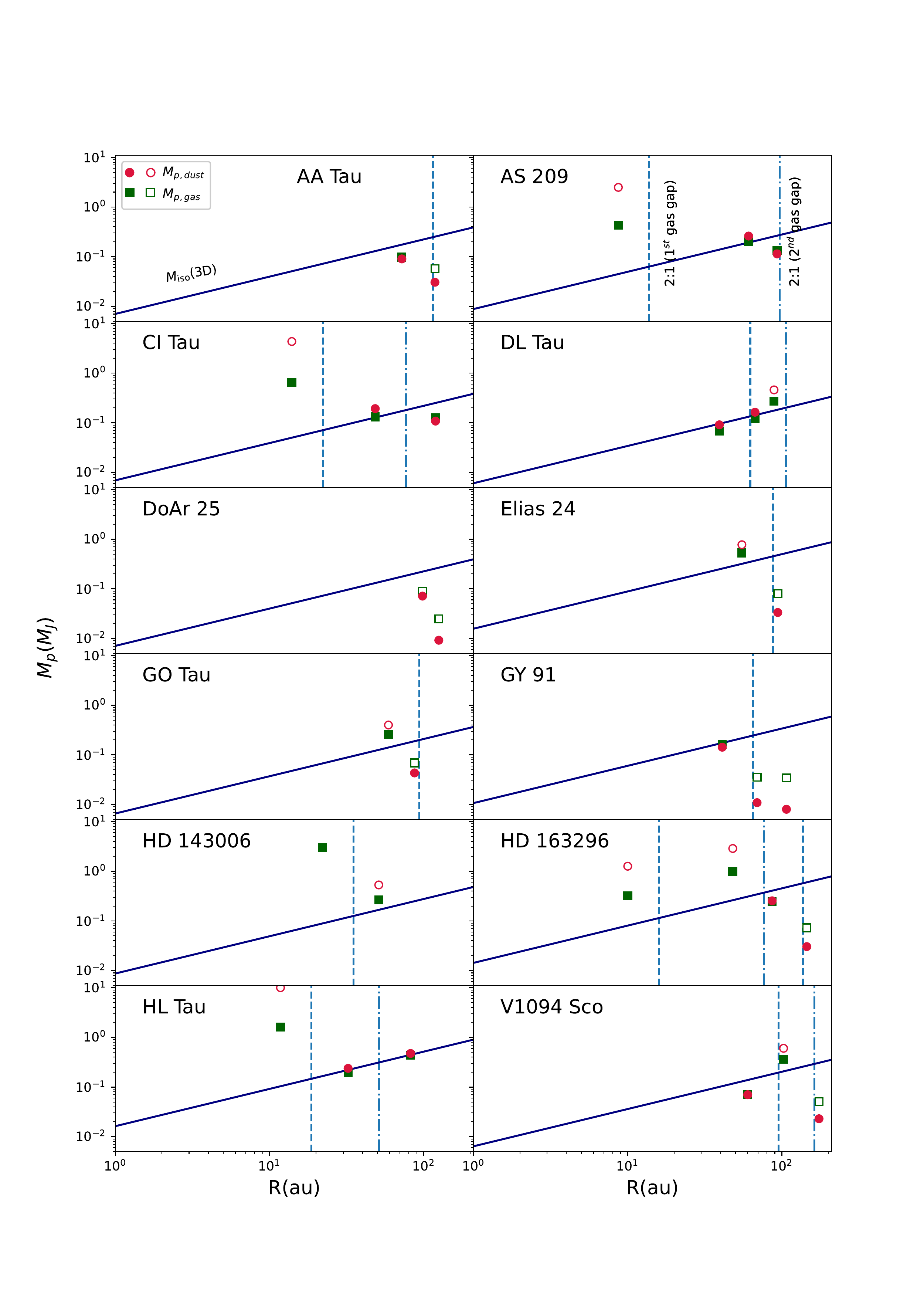}
	\caption{Configuration of the predicted planetary systems of the multi-gap disks ($ \alpha =\num{e-3}$). The solid and dashed blue line shows the pebble isolation mass obtained from 3D simulations \citep{Bitsch2018}. If the predicted mass does not satisfy the corresponding criterion, an empty marker is used; otherwise, a filled marker is used. The dashed vertical lines indicate the position of the outer 2:1 period ratio for inner planets that have opened gas gaps. We use two line styles to denote the 2:1 lines from different planets.}
	\label{fig:mass_vs_R_expo}
\end{figure*}

We estimate planetary masses in the observed disk gaps according to our gap classification, assuming that planets are fully responsible for creating the gap substructures, and gaps and planets are in one-to-one
correspondence. The mass prediction result is summarized in Table~\ref{tab:planetary_mass}. In the case of group III, we list both $ M_{\rm p, dust} $ and $ M_{\rm p, gas} $.

\subsubsection{Case with fiducial $ \alpha =\num{e-3} $}

The predicted planetary mass ranges from $ \SI{1.6}{M_\oplus} $ to $ \SI{4.8}{M_J} $, but with significant diversities. Figure \ref{fig:comp_obs_mvsr} plots the distribution of the planets, $ M_p/M_* $ against the semi-major axis for $\alpha = \num{e-3}$.  We also plot the observed exoplanet population (extracted from \url{exoplanet.eu}, as of March 2021) with confirmed planetary mass, semi-major axis and stellar mass for comparison. We specifically label several widely-separated giant planets observed by direct imaging including HR 8799, $ \beta $ Pictoris and PDS 70, which would be relevant to compare with our prediction.

Figure \ref{fig:comp_obs_mvsr} indicates that our predicted planets are located in a fairly distinct region relative to that of the observed population. It may imply that these observed planets have experienced significant inward orbital migration before the disk dispersal due to disk-planet interaction \citep{Zhang2018,Lodato2019}. Furthermore, their mass should have increased via accretion during the migration. A possible model of such evolutionary tracks is introduced by \cite{Tanaka2019}, which suggests the following relation between mass and semi-major axis:
\begin{equation}\label{eqn:evo_track}
  \ln\left(\frac{R_p}{\SI{1}{au}}\right)
  + \frac{2}{3}\left(\frac{M_p}{0.011M_*}\right)^{2/3} = \mathcal{C},
\end{equation}
where $ \mathcal{C} $ is a constant. Just for reference, we plot examples of the tracks with the range of $\mathcal{C}$ values that match our predicted planets. While a fraction of observed planets is consistent with the evolutionary tracks, a majority of them is far outside of the region that the simply model predicts. At this point, the reason of the discrepancy is not clear, and more detailed study of their evolution is necessary, which we plan to show elsewhere.

The architecture of the planets predicted from the multi-gap systems is shown in Figure \ref{fig:mass_vs_R_expo}. Mass predictions based on $ M_{\rm p, dust} $ and $ M_{\rm p, gas} $ are plotted in circles and squares, respectively. The filled (open) symbols indicate that they are consistent (inconsistent) with the criteria C1 and C2. The pebble isolation mass is also indicated by a straight line. As mentioned before, the planetary mass of most disks generally decreases against $ R $, except the cases of DL Tau, HD 163296, HL Tau and V1094 SCO: the masses of the three planets in the DL Tau system increase with $ R $; for the rest of the systems, there exist at least one pair of planets that does not follow the trend. We found planets that violate this trend are always gas-gap planets. If an outer planet forms first and later approaches the inner planet via inward migration, it may start runaway gas accretion earlier than the inner planet and quench its growth by shielding the gas flow. This implies that the architecture of a multi-planetary system is sensitive to its initial configuration.

Figure \ref{fig:mass_vs_R_expo} also plots the locations of the outer 2:1 resonance period for the inside gas-gap planets (including marginal ones in group III) in dashed vertical lines. In many systems (i.e. AA Tau, AS 209, GY 91, Elias 24, DL Tau, GO Tau, HD 163296 and V1094 SCO) the gas-gap planet has a dust-gap planet close to its 2:1 period ratio. It may be due to the fact that the dust-gap planet is likely to form at the 2:1 resonance place of the gas-gap (or relatively large) planet. Alternatively, such a configuration can be the consequence of the combination of planetary migration and resonance trap. These mass orders may be related to alternative formation scenarios of planets, which will be discussed in the next section.

\subsubsection{$ \alpha $ dependence}
\label{sec:alpha_dependence}

So far we have adopted the fiducial $ \alpha $ value \num{e-3}. In fact, however, both $ M_{\rm p, gas} $ and $ M_{\rm iso} $ depend on $ \alpha $, and different values of $ \alpha $ affect both classification criteria and the predicted mass. To investigate the $ \alpha $ dependence, we calculate the planetary masses when $ \alpha = \num{e-4}$ and $\num{e-2} $ in addition to the fiducial value and list the results in Table \ref{tab:planetary_mass}. We also present the gap classification results in Appendix \ref{sec:appendix_class_gaps_appendix}. When the $\alpha$ is lower, C2 is less likely to be satisfied as $ M_{\rm p,gas} $ is scaled by $ \alpha^{1/2} $, so more gaps are classified as dust-only gaps (group I). As $ \alpha $ increases, in contrast, $ M_{\rm p, gas}$ increases and $ M_{\rm p, dust}/M_{\rm iso}$ decreases, therefore both C1 and C2 are easier to be satisfied, and the number of indistinguishable gaps (group III) increases.

\begin{figure}
\centering 
\includegraphics[width=\linewidth]{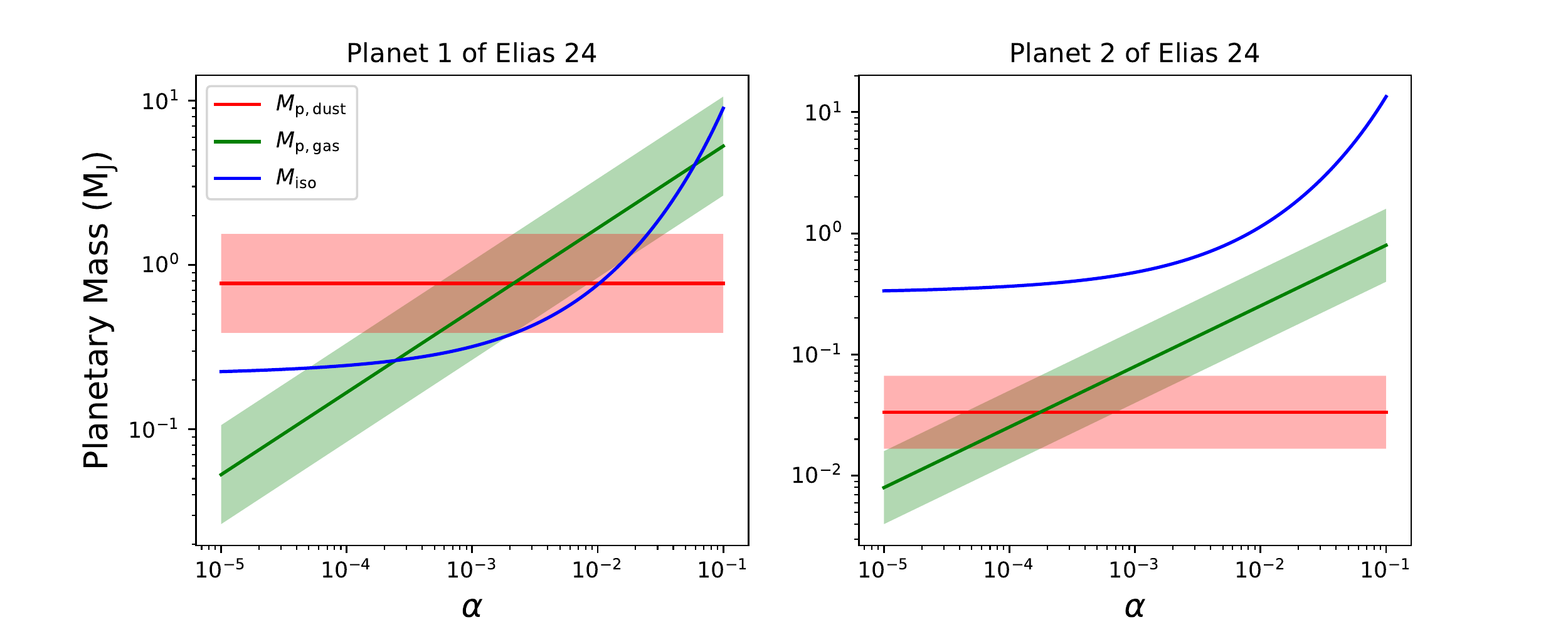} 
\caption{$ M_{\rm p, dust} $ (red line), $ M_{\rm iso} $ (blue line) and $ M_{\rm p, gas} $ (green line) against $ \alpha $ in Elias 24 disk. The shaded region covers a factor of two uncertainty of the mass prediction.} 
\label{fig:mvsalpha_elias24}
\end{figure}

To illustrate the $\alpha$-dependence, we choose Elias 24 as an example. Figure \ref{fig:mvsalpha_elias24} shows how the mass changes for the gaps in Elias 24 system by varying $\alpha$. For gap 1 (innermost gap), when $ \num{5e-5} <\alpha < \num{2e-3} $, $ M_{\rm p, dust}/2 <  M_{\rm iso}$ is not satisfied, while $ 2M_{\rm p, gas} >  M_{\rm iso}$ is satisfied. Therefore, only the gas gap interpretation is valid and the gap is in group II. In the range $ \num{2e-3} \leq \alpha$, both dust gap and gas gap interpretations are valid, therefore the gap is in group III. However for gap 2 (outer gap), the condition $ M_{\rm iso} > 2M_{\rm p, gas} $ holds for all range of $ \alpha $, therefore the gap cannot be interpreted as a gas gap and is always in group I. Note that the original viscosity range in \cite{Bitsch2018} is $ \num{e-4} \leq \alpha \leq \num{e-2} $. Outside of this range, the prediction of $ M_{\rm iso} $ becomes an extrapolation of \cite{Bitsch2018} that may be not accurate.

We note that while $M_{\rm p, dust}$ defined by equation~(\ref{eqn:pltmass_Hill}) does not exhibit any explicit dependence on $\alpha$, it should depend on $\alpha$ in reality. We discuss this issue in Section~\ref{sec:discussion}.

\begin{table*}[!ht]
	\centering
	\caption{Planetary mass estimated from the gaps}
	\resizebox{\linewidth}{!}{
		\begin{threeparttable}
			\begin{tabular}{lcccccccccc}
				\toprule 
		          	\multirow{2}{*}{Disk} &    \multirow{2}{*}{Gap index}   & \multicolumn{3}{c}{$\alpha = \num{e-2}$} & \multicolumn{3}{c}{$\alpha = \num{e-3}$} & \multicolumn{3}{c}{$\alpha = \num{e-4}$} \\
			     &  & Group          & $ M_{\rm p,dust} $($ \si{M_J} $) & $ M_{\rm p,gas} $($ \si{M_J} $)                 & Group            & $ M_{\rm p,dust} $($ \si{M_J} $) & $ M_{\rm p,gas} $($ \si{M_J} $)                 & Group            & $ M_{\rm p,dust} $($ \si{M_J} $) & $ M_{\rm p,gas} $($ \si{M_J} $)                   \\
				\hline
				\textit{Single-gap systems}\\
			DM Tau          & 1 & III & 0.520 & 1.052  & II  & -     & 0.333 & II & -     & 0.105 \\
			DN Tau          & 1 & I   & 0.005 & -      & I   & 0.005 & -     & I  & 0.005 & -     \\
			DS Tau          & 1 & II  & -     & 3.429  & II  & -     & 1.084 & II & -     & 0.343 \\
			Elias 20        & 1 & I   & 0.025 & -      & I   & 0.025 & -     & I  & 0.025 & -     \\
			Elias 27        & 1 & III & 0.082 & 0.353  & I   & 0.082 & -     & I  & 0.082 & -     \\
			FT Tau          & 1 & III & 0.048 & 0.138  & III & 0.048 & 0.044 & I  & 0.048 & -     \\
			GW Lup          & 1 & I   & 0.037 & -      & I   & 0.037 & -     & I  & 0.037 & -     \\
			HD 135344B      & 1 & I   & 0.016 & -      & I   & 0.016 & -     & I  & 0.016 & -     \\
			HD 142666       & 1 & III & 0.313 & 0.470  & II  & -     & 0.149 & II & -     & 0.047 \\
			HD 169142       & 1 & II  & -     & 3.557  & II  & -     & 1.125 & II & -     & 0.356 \\
			HD 97048        & 1 & II  & -     & 15.062 & II  & -     & 4.763 & II & -     & 1.506 \\
			IM Lup          & 1 & I   & 0.041 & -      & I   & 0.041 & -     & I  & 0.041 & -     \\
			IQ Tau          & 1 & III & 0.045 & 0.148  & III & 0.045 & 0.047 & I  & 0.045 & -     \\
			MWC 480         & 1 & II  & -     & 4.241  & II  & -     & 1.341 & II & -     & 0.424 \\
			RU Lup          & 1 & III & 0.044 & 0.165  & I   & 0.044 & -     & I  & 0.044 & -     \\
			RXJ 1615.3-3255 & 1 & III & 0.614 & 1.360  & II  & -     & 0.430 & II & -     & 0.136 \\
			RY Tau          & 1 & I   & 0.054 & -      & I   & 0.054 & -     & I  & 0.054 & -     \\
			SR 4            & 1 & II  & -     & 1.415  & II  & -     & 0.447 & II & -     & 0.141 \\
			Sz 114          & 1 & I   & 0.005 & -      & I   & 0.005 & -     & I  & 0.005 & -     \\
			Sz 129          & 1 & I   & 0.016 & -      & I   & 0.016 & -     & I  & 0.016 & -     \\
			Sz 98           & 1 & III & 0.134 & 0.490  & III & 0.134 & 0.155 & I  & 0.134 & -     \\
			UZ Tau E        & 1 & I   & 0.009 & -      & I   & 0.009 & -     & I  & 0.009 & -     \\
			V1247 Ori       & 1 & I   & 0.100 & -      & I   & 0.100 & -     & I  & 0.100 & -     \\
			\hline
			\textit{Multi-gap systems}\\
			AA Tau          & 1 & III & 0.090 & 0.310  & III & 0.090 & 0.098 & I  & 0.090 & -     \\
			AA Tau          & 2 & I   & 0.031 & -      & I   & 0.031 & -     & I  & 0.031 & -     \\
			DoAr 25         & 1 & III & 0.071 & 0.280  & I   & 0.071 & -     & I  & 0.071 & -     \\
			DoAr 25         & 2 & I   & 0.009 & -      & I   & 0.009 & -     & I  & 0.009 & -     \\
			Elias 24        & 1 & III & 0.772 & 1.673  & II  & -     & 0.529 & II & -     & 0.167 \\
			Elias 24        & 2 & I   & 0.033 & -      & I   & 0.033 & -     & I  & 0.033 & -     \\
			GO Tau          & 1 & III & 0.399 & 0.827  & II  & -     & 0.262 & II & -     & 0.083 \\
			GO Tau          & 2 & I   & 0.043 & -      & I   & 0.043 & -     & I  & 0.043 & -     \\
			HD 143006       & 1 & II  & -     & 9.415  & II  & -     & 2.977 & II & -     & 0.942 \\
			HD 143006       & 2 & III & 0.531 & 0.836  & II  & -     & 0.264 & II & -     & 0.084 \\
			AS 209          & 1 & II  & -     & 1.373  & II  & -     & 0.434 & II & -     & 0.137 \\
			AS 209          & 2 & III & 0.260 & 0.632  & III & 0.260 & 0.200 & I  & 0.260 & -     \\
			AS 209          & 3 & III & 0.114 & 0.428  & III & 0.114 & 0.135 & I  & 0.114 & -     \\
			CI Tau          & 1 & II  & -     & 2.075  & II  & -     & 0.656 & II & -     & 0.208 \\
			CI Tau          & 2 & III & 0.192 & 0.416  & III & 0.192 & 0.132 & I  & 0.192 & -     \\
			CI Tau          & 3 & III & 0.108 & 0.397  & III & 0.108 & 0.125 & I  & 0.108 & -     \\
			DL Tau          & 1 & III & 0.091 & 0.215  & III & 0.091 & 0.068 & I  & 0.091 & -     \\
			DL Tau          & 2 & III & 0.163 & 0.387  & III & 0.163 & 0.122 & I  & 0.163 & -     \\
			DL Tau          & 3 & III & 0.459 & 0.858  & II  & -     & 0.271 & II & -     & 0.086 \\
			GY 91           & 1 & III & 0.143 & 0.515  & III & 0.143 & 0.163 & I  & 0.143 & -     \\
			GY 91           & 2 & I   & 0.011 & -      & I   & 0.011 & -     & I  & 0.011 & -     \\
			GY 91           & 3 & I   & 0.008 & -      & I   & 0.008 & -     & I  & 0.008 & -     \\
			HL Tau          & 1 & II  & -     & 5.116  & II  & -     & 1.618 & II & -     & 0.512 \\
			HL Tau          & 2 & III & 0.238 & 0.618  & III & 0.238 & 0.195 & I  & 0.238 & -     \\
			HL Tau          & 3 & III & 0.474 & 1.389  & III & 0.474 & 0.439 & I  & 0.474 & -     \\
			V1094 Sco       & 1 & III & 0.070 & 0.225  & III & 0.070 & 0.071 & I  & 0.070 & -     \\
			V1094 Sco       & 2 & III & 0.602 & 1.153  & II  & -     & 0.365 & II & -     & 0.115 \\
			V1094 Sco       & 3 & I   & 0.023 & -      & I   & 0.023 & -     & I  & 0.023 & -     \\
			HD 163296       & 1 & II  & -     & 1.010  & II  & -     & 0.320 & II & -     & 0.101 \\
			HD 163296       & 2 & II  & -     & 3.147  & II  & -     & 0.995 & II & -     & 0.315 \\
			HD 163296       & 3 & III & 0.254 & 0.779  & III & 0.254 & 0.246 & I  & 0.254 & -     \\
			HD 163296       & 4 & I   & 0.030 & -      & I   & 0.030 & -     & I  & 0.030 & -    \\
							\hline
			\end{tabular}
		\end{threeparttable}
	}
	\label{tab:planetary_mass}
\end{table*}

\section{Discussion}
\label{sec:discussion}
\subsection{Comparison with previous studies}
\label{sec:comparison}
Our sample disks are largely identical to those adopted by \cite{Zhang2018} and \cite{Lodato2019}, and thus we compare the results to see the agreement. \cite{Zhang2018} carried out gas-dust two-fluid hydrodynamical simulations and developed a more sophisticated method to derive the gas gap width by considering three different dust models (DSD1, `1mm', DSD2). Compared with their study, our study simply assumes that the gas gap width matches that of the dust gap due to strong gas-dust coupling, which is not necessarily valid if the coupling between dust and gas is weak. In principle, if the dust grains are decoupled from the gas \citep{Zhu2012,Weber2018}, the dust gap should be wider than its associated gas gap. Therefore, the planetary masses predicted by the well-coupled DSD1 model of \cite{Zhang2018} are supposed to be comparable to ours, while those predicted by the less-coupled DSD2 model should be smaller.

At $ \alpha =\num{e-3}$, the majority of the predictions given by DSD1 model from \cite{Zhang2018} are similar to ours within a factor of two (it is the same level of the uncertainly that we used), such as \nth{1} planet of Elias 24 ($ 45 \% $), Elias 27 ($ 31 \% $) and GW Lup ($ 21 \% $). However, \cite{Zhang2018} gave significant larger predictions than ours for the inner or wide gaps: for instance, the predicted masses of the innermost gaps in HD 143006 and SR 4 differ by nearly a factor of five, and even the masses given by the `DSD2' model are larger than ours. This discrepancy may be attributed to the difference in measurement of the gap width and computation time between \cite{Kanagawa2016} and \cite{Zhang2018}. The predictions of \cite{Zhang2018} are based on fitting of gaps at \num{1000} planetary orbits, while the result of \cite{Kanagawa2016} is based on stationary shape of gaps in long-term simulations. Also the computational time required to reach the stationary shape is longer for a wider gap (see Figure 14 of \cite{Kanagawa2017}). If the gap width does not reach the stationary value within \num{1000} orbits, \cite{Zhang2018} may give a larger planet mass as compared with our predictions. In the case of the innermost gap of HD 143006, the stellar age ($\sim$\SI{4}{Myr}) corresponds to $\sim$\num{55000} orbits at the location of the innermost gap (\SI{21}{au}), and hence the simulation time could be insufficient for the gap to reach its stationary width.

We also compare our results with those of \cite{Lodato2019}. When a gap is classified as dust-only gaps (group I), we adopt the same Hill radius scaling relation as \cite{Lodato2019}, and thus our predicted planetary mass for group I gaps most agree with those from \cite{Lodato2019} within $ 50\% $ uncertainty level, such as RY Tau ($ 35\% $) and the \nth{2} planet of GO Tau ($ 48\% $). This deviation is mainly due to the different stellar masses and gap widths adopted. Nevertheless, the planetary mass predicted from group II gaps in our study is significantly smaller than that in their study. For example, \cite{Lodato2019} predicted mass of the first gap of CI Tau to be $ \SI{15.7}{M_J} $, while the same planet is predicted to be $ \SI{0.85}{M_J} $ in our fiducial case with $ \alpha = \num{e-3} $. Apart from the different gap width data we adopted, it is because $ M_{\rm p,gas} $ given by equation (\ref{eqn:pltmass_Kanagawa}) scales with $ \Delta_{\rm gap}^2 $, while $ M_{\rm p,dust} $ based on the Hill radius relation scales with $ \Delta_{\rm gap}^3 $ (also see Appendix \ref{sec:appendix}). Physically, it means that in the gas gap regime the dominant planet-gas interaction can create a wider gap than that expected by the Hill radius argument. Therefore, the planetary mass in this regime may be overestimated if equation (\ref{eqn:pltmass_Hill}) is used. It should be noted that in most cases, the width of the gas gap given by equation~(\ref{eqn:pltmass_Kanagawa}) is wider than that of the dust gap predicted by equation~(\ref{eqn:pltmass_Hill}).

We would like to point out that theoretically, it is possible that the dust gap is wider than the gas gap; the dust gap width, $\Delta_{\rm gap} $ in equation (\ref{eqn:pltmass_Hill}), is independent of the gas parameters, while the gas gap width, $\Delta_{\rm gap}$ in equation (\ref{eqn:pltmass_Kanagawa}), is scaled with viscosity as $\alpha^{-1/4}$, and with scale height as $(h/R)^{-3/4} $. Therefore, for a given planetary mass $M_{p} > M_{\rm iso} $, it is always possible to find a set of gas disk parameters (i.e. large enough $ \alpha $, $ h/R $) with which the dust gap width by equation (\ref{eqn:pltmass_Hill}) is wider than that of the gas gap by equation (\ref{eqn:pltmass_Kanagawa}). In such cases, $M_{\rm p,dust}$ is larger than $M_{\rm iso}$, but the edge of the dust gap may be apart from the gas pressure bump. Thus it might be appropriate to still use equation~(\ref{eqn:pltmass_Hill}) to estimate the planet mass.

Having said so, however, the parameter ranges for such cases (roughly $ \alpha > 0.01$ and $ h/R > 0.1 $) are fairly different from those commonly adopted by previous studies. Moreover, the dust grains have to be decoupled from the gas; if the dust is strongly coupled to the gas, the width of the dust gap should well match that of the gas gap. In this case, the dust gap width expected by equation~(\ref{eqn:pltmass_Hill}) is an overestimate. In the intermediate regime between the strongly-coupled and the decoupled cases, the dust gap width roughly agrees with the gas gap width, and equation~(\ref{eqn:pltmass_Kanagawa}) is still a good approximation in the gas gap regime.

\subsection{Implications on planetary formation}
\label{sec:implications}
Our classification of the observed gaps has shown gaps at inner region of the disk are mostly gas gaps (group II), while those at the outer region are mostly dust gaps (group I). The cores at the inner region of the disk undergo faster accretion due to high surface densities of both pebbles and gas, so they more likely become gas giants that can open wider gas gaps. On the other hand, planets at outer region of the disk undergo slow accretion because of insufficient material, and runaway gas accretion is also harder to be triggered because the larger aspect ratio leads to larger pebble isolation mass. In effect, it is harder for planets to open gaps in both gas and dust profiles at the outer region.

The abundance of dust gaps at the outer region also implies that the outer planets are in fact small in general, suggesting these distant planets are not those formed via gravitational instability, which should be much more massive and open gas gaps instead \citep{Mayer2002}. However, it remains unclear that how the cores of the outer planets form at such a large distance to the star, since the formation of the solid cores via collision at the outer region is very inefficient due to low density and temperature. It is also possible that cores are not formed \textit{in situ}: they may form at the inner dust-rich region first, then migrate outwards and inward again after growing up \citep[e.g.][]{Paardekooper2005,Paardekooper2011a}.

The configurations of the multi-planetary systems shown by Figure \ref{fig:mass_vs_R_expo} suggest in general the planetary mass decreases with the radius. In a single-planetary systems, the planet can migrate freely inward and grow quickly at the inner region. In multi-planetary systems, if a gas giant is formed first at the interior, the strong 2:1 orbital resonance will prevent the outer planet from migrating further inwards, therefore in this case the inner planet always grows faster than the outer planets and the mass order is retained. 

However, there are four disks out of this trend, namely DL Tau, HD 163296, HL Tau and V1094 SCO. For these four disks, it is possible that the outer planet is initially larger than the inner planet. As a result, it migrates to the current position and starts gas accretion earlier, which shields the inner planets from accessing accretion materials and therefore quenches its growth (also see \cite{Wang2020}). Therefore, the fate of an individual planet strongly depends on its birth place and time relative to other planets. While investigating the evolution of a multi-planetary system, it is important to consider the coupling between planets, as the planet-planet interaction can effectively shape the orbital and mass configuration.

We also found that in half of our sample disks, the outermost gas-gap is accompanied with a dust-gap planet close to its outer 2:1 mean motion resonance. The resonant trap is one of the possible explanations: the dust-gap planet migrates inwards faster via type I migration, while the migration speed of the inner gas-gap planet is slower due to the gap-opening effect \citep{Kanagawa2018}. Once the outer planet approaches the inner planet, it will be captured due to resonance and start co-migration \citep[e.g.][]{Mustill2011,Deck2015}. An alternative explanation is that the density of large pebbles or planetesimals may be enriched at the resonance sites \citep[e.g.][]{Wyatt2003}, therefore a planet is preferred to form near the resonance. Further investigations are required to discuss the possibilities of the respective scenario.

Finally, the evolution tracks on Figure \ref{fig:comp_obs_mvsr} do not pass through the region of the observed hot Jupiters. This implies that those gas-gap opening planets in our predicted population are unlikely to be the origin of the hot Jupiters, at least before the dispersal of the disk. It is still possible for low mass planets to become hot Jupiters, but it depends on whether there exist a inner gas giant in the disk that can stop the inward migration of low mass planet via resonance capture. Alternatively, if the configuration of the planetary system is dynamically unstable, hot Jupiters can be formed via later stage scattering followed by tidal circularization \citep[e.g.][]{Nagasawa2008}. Therefore, the stability of the planetary systems is crucially important to determine the final configurations, and that will be covered in our second paper.

\subsection{Caveats}
There are a few caveats while applying our model. Firstly, we assume that there is a one-to-one correspondence between the gap and the planet, which may not be true for all the gaps. It is possible that two close-in planets can create a wide common gap \citep[e.g.][]{Cimerman2018} that can be misinterpreted as a single gap created by a massive gas giant. However, the possibility for the common gap interpretation may be restricted to the wide gaps with high $ \Delta_{\rm gap}/R_{\rm gap} $, such as the first gaps of DS Tau and HD 143006 systems. For those narrow gaps with $ \Delta_{\rm gap}/R_{\rm gap} \sim 0.1$ in our sample, they are unlikely to be common gaps, because otherwise each planet is too small to create even a dust gap. Moreover, a single planet can generate multiple gaps and rings in a very low viscosity disk \citep{Dong2017,Bae2017}, meaning we may overestimate the number of planets. We need to have better knowledge about the gap structure in these contexts and more observational details about the gap morphology to identify these cases.

Secondly, while applying equation~(\ref{eqn:pltmass_Kanagawa}), we always assume the perfect dust-gas coupling by equating the width of the dust gap to that of the gas gap. The degree of gas-dust coupling depends on the Stokes number, which is a function of the dust grain size and gas density \citep[e.g.][]{Takeuchi2005}. In the well-coupled situation, our results are similar to \cite{Zhang2018}; nevertheless, when the dust is strongly decoupled from the gas, our prediction may overestimate the planetary mass, even though we have considered a relatively large uncertainty range in our estimate.

Thirdly, though we neglect $\alpha$-dependence in $M_{\rm p, dust}$ given by equation~(\ref{eqn:pltmass_Hill}), $M_{\rm p, dust}$ could depend on $\alpha$ through the coupling between the gas and dust grains.  In the strong coupling limit, the dust particles would depend on $\alpha$ similar as the gas, {\it i.e.,} $\propto \alpha^{1/4}$ indicated by equation~(\ref{eqn:pltmass_Kanagawa}). If the dust grains are only weakly coupled to the gas, on the other hand, the $\alpha $-dependence should vanish for large St (in the limit $ \alpha $/St$ \rightarrow 0$) as indicated by \cite{Dipierro2017}.

The dust particles visible in the ALMA band fall within these two limits, and therefore the $\alpha $-dependence of $ M_{\rm p, dust}$ should be small and does change our current result.  Moreover, $M_{\rm p, dust}$ should depend on the Stokes number \text{St} as well as $\alpha$.  In a weakly coupled dust regime, \cite{Dipierro2017} suggest that the dust gap width only weakly depends on $ \text{St}^{1/4} $, so the planetary mass can still be fairly well constrained even if an order of magnitude variance of St is taken into account.  For the intermediate regime, the limitation of adopting equation~(\ref{eqn:pltmass_Hill}) may be kept in mind, and we note here that a more robust prediction of the planetary mass should consider both $\alpha$ dependence and St estimation carefully.

Moreover, our classification of gap depends on the pebble isolation mass that is sensitive to the aspect ratio, so the grouping result is subject to change when the temperature profile deviates from that given by equation (\ref{eqn:dust_temp_fit}). Apart from the aforementioned factors, the predicted planetary mass also depends on properties of the dust grains, gas density, viscosity, and the way to measure gap widths. Hence, we should stress that our predicted mass can deviate by a factor of two, or larger in some extreme cases. Our results can be improved if we can more reliably obtain the width of the gas gap with considerations of detailed dust models, temperature profile and gas density profile.
\section{Summary and conclusion}
\label{sec:summary}
The gap structures have been identified on many of the observed PPDs, and planets are the most conventional interpretation of their origin. The scaling relations found by previous studies can be used to deduce the planetary mass from the shape of the gap. In order to make appropriate predictions, it is necessary to know whether the gap exists in both the gas/dust profiles or only exist in the dust profile. In this paper, we purpose a method of solving this degeneracy by considering the core accretion scenario of planetary formation. Since the pebble isolation mass is always associated with a deep gas gap, if the predicted planetary mass is larger than pebble isolation mass, it can be regarded as a gap in both gas/dust profiles. Otherwise, the gap is considered as only a dust gap. By translating these mass criteria to gap classification criteria using two mass scaling relations, we can classify the observed gaps into four groups, and then predict the planetary mass according to the grouping. Based on our analysis of 35 disks with 55 gaps, our main results are:
\begin{enumerate}
	\item For our fiducial value of $ \alpha $(= $\num{e-3}$),  $ 21 $ out of $ 55 $ gaps are classified as dust-only gaps, $ 19 $ are gas gaps, and $ 15 $ remain indistinguishable (table \ref{tab:gap_classification}). Our criteria can distinguish between dust-only and dust-gas gaps for the majority of the observed gaps ($ 73\% $). The viscosity $ \alpha $ can significantly change the interpretation of the gaps, as discussed in section \ref{sec:alpha_dependence}. As $ \alpha $ increases, the number of indistinguishable gaps also increases. We show a particular example of $ \alpha $ dependence in Figure \ref{fig:mvsalpha_elias24} and the predictions are summarized in table \ref{tab:planetary_mass}.
	
	\item Distribution of gaps in multi-gap systems shows that inner gaps are mostly gas gaps, while the outer gaps are mostly dust-only gaps. The predicted planetary mass in general decreases with increasing orbital radius. As discussed in section \ref{sec:implications}, this trend can be explained by the difference of the accretion rates and pebble isolation mass at inner and outer disk regions, and is consistent with the core accretion scenario.
	
	\item The predicted planetary mass ranges from $ \SI{1.6}{M_\oplus} $ to $ \SI{4.8}{M_J} $. The evolution tracks for single planets suggest our predicted planets have potential to be the origin of the widely separated giant planets, such as planets in HR 8799 and PDS 70 (Figure \ref{fig:comp_obs_mvsr}).  We also found that observed hot Jupiters are not located on the evolution tracks associated with planets that we predict. As discussed in section \ref{sec:implications}, it may imply that the dynamical instability is required to form hot Jupiters. Alternatively, their origin is different from those gap opening planets observed in PPDs. 
	
	\item Closely outside of the 2:1 resonance line of the inner planet that opens gas gap, there often exist a small planet that opens only a dust gap (Figure \ref{fig:mass_vs_R_expo}). As discussed in section \ref{sec:implications}, it can be related to the resonance capture due to convergent migration. Alternatively, the orbital resonance may be causally related to the planetary core formation.
\end{enumerate}
Our results suggest that the core accretion scenario is consistent with the the observed gap structure, and the planet-planet interaction in the multi-planetary systems plays an important role to shape their architecture. In our Paper II, we will focus on multi-planetary systems and carry out $ N $-body simulations based on the initial conditions that we have obtained. Before the dispersal of the disk, we will implement both migration and accretion with different disk parameters. After the disk dispersal, we will continue integrating the systems with gravitational force to \si{Gyr} timescale to assess the dynamical stability of the configuration. Eventually we will compare our synthesized systems with the observed exoplanetary population and see to what extent we can connect the observed disks to the diversity of the observed exoplanets.
\section*{Acknowledgment}
We thank an anonymous referee for numerous constructive criticisms that we believe significantly improved the earlier manuscript of the present paper. We also thank Ruobing Dong for clarifying the meaning of the dust temperature defined in \cite{Dong_Najita2018}. S.W. acknowledges support from the Chinese Scholarship Council (CSC). K.D.K. acknowledges support from the Research Center for the Early Universe, the University of Tokyo. This work is supported partly by the Japan Society for the Promotion of Science (JSPS) Core-to-Core Program “International Network of Planetary Sciences”, and also by JSPS KAKENHI grant No. JP18H01247 and JP19H01947 (Y.S.), and JP19K14779 (K.D.K).

\appendix
\section{Interpretation of gap classifications at fixed locations}
\label{sec:appendix}
\begin{figure}[h!]
	\centering
	\gridline{\fig{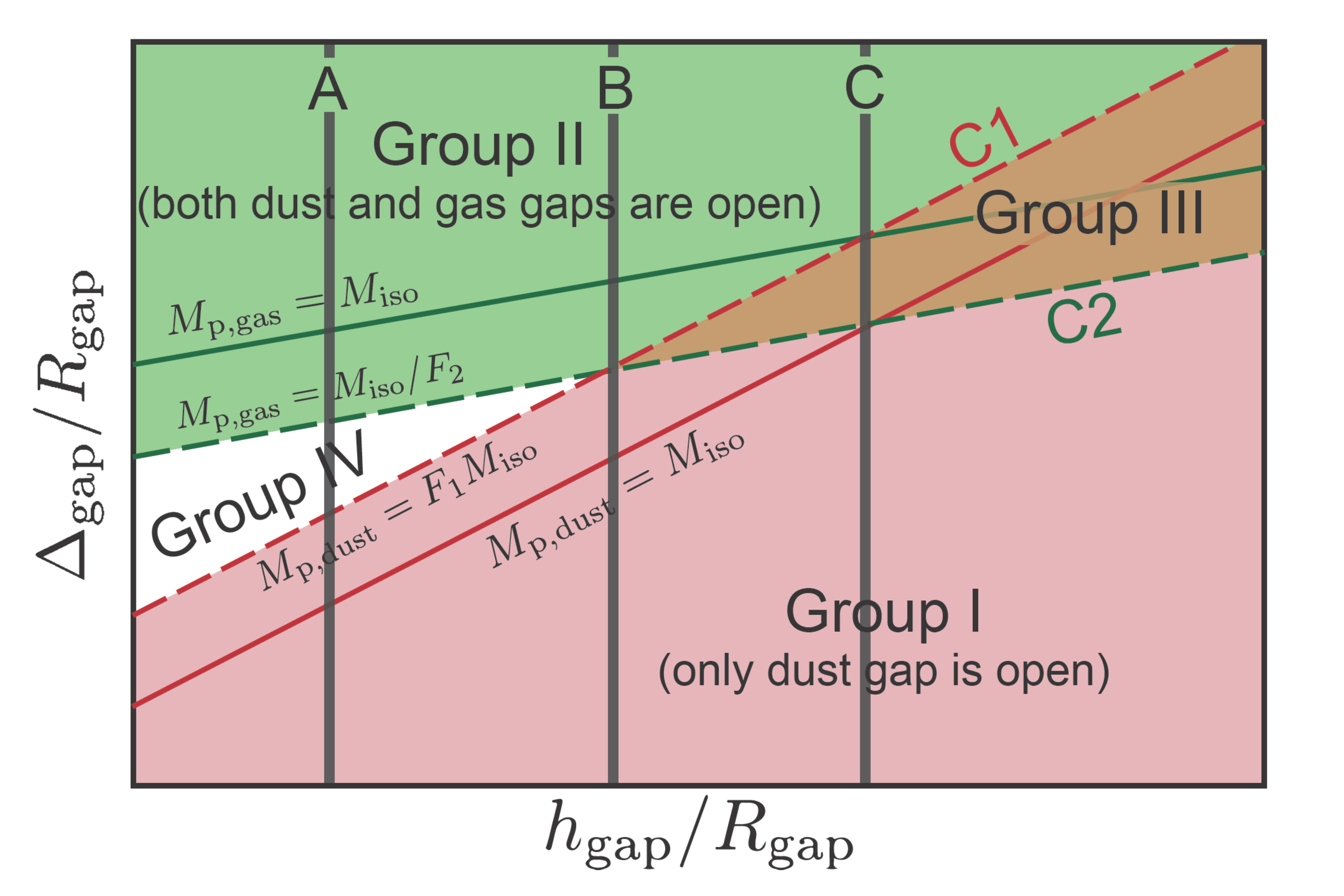}{0.70\linewidth}{(a)}}
	\gridline{\fig{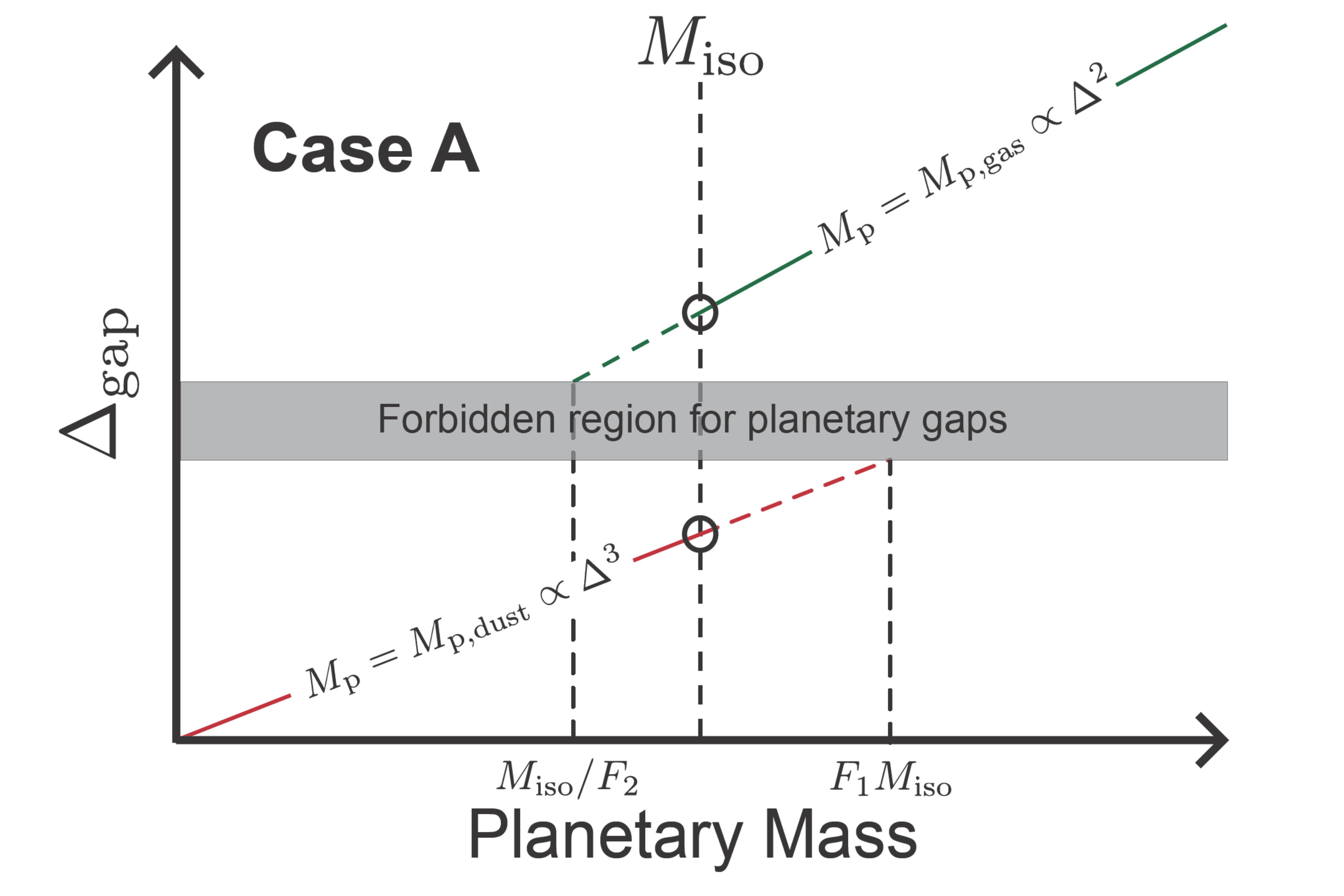}{0.5\linewidth}{(b)}
			  \fig{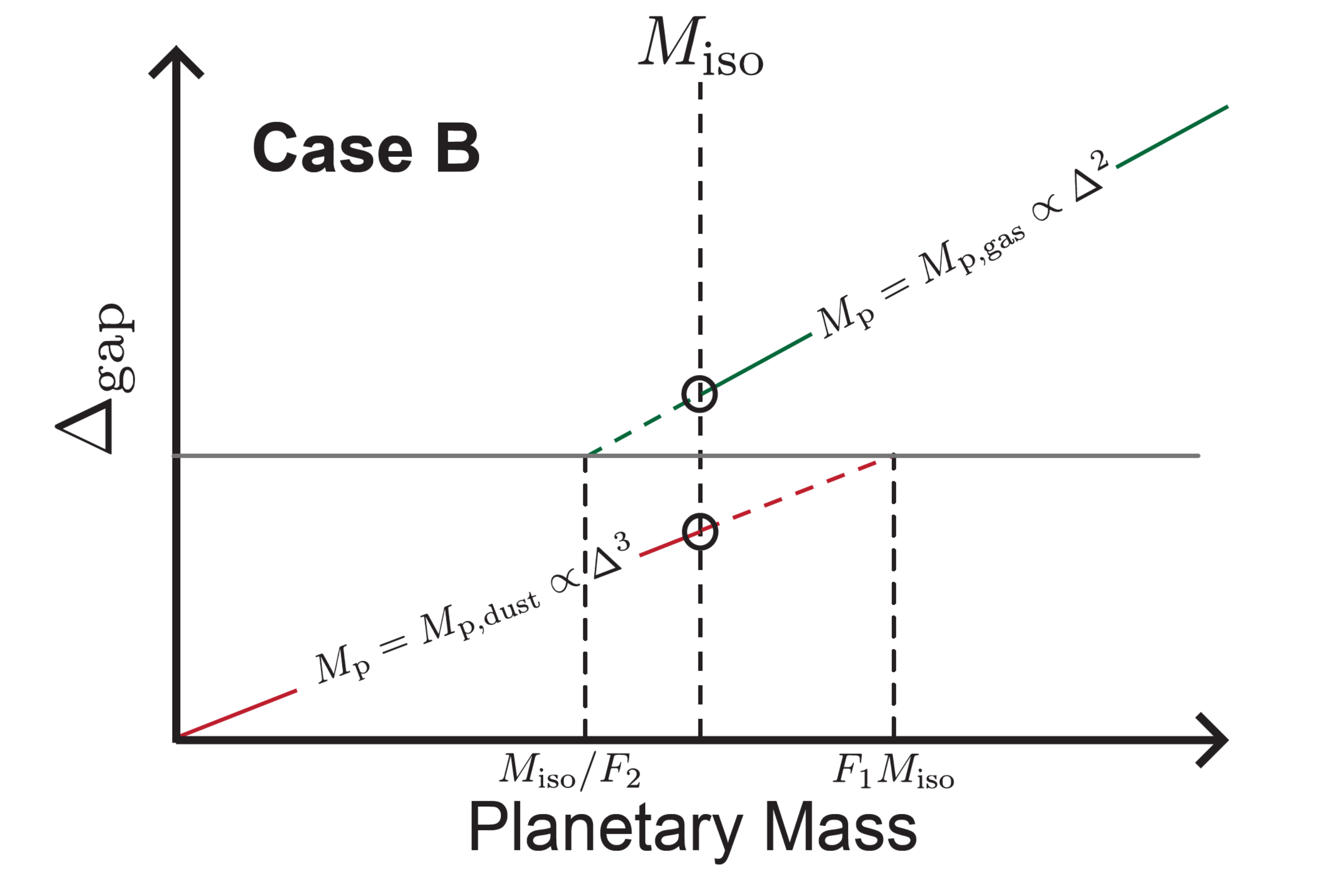}{0.5\linewidth}{(c)}}
  	\gridline{\fig{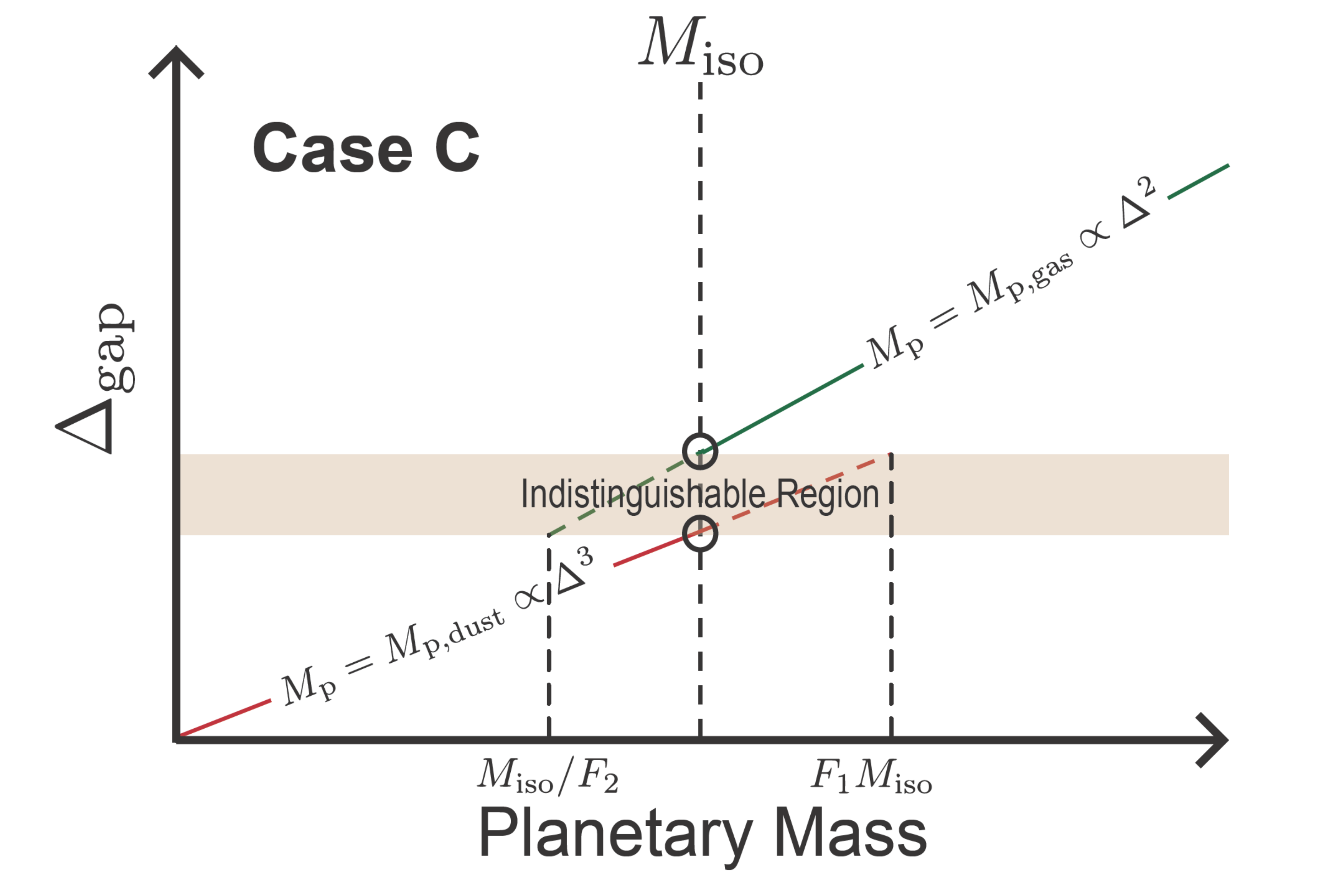}{0.5\linewidth}{(d)}}
  	\caption{(a) Same as figure \ref{fig:gpillustration} but with A, B, C lines indicting three fixed locations at the same disk. (b)(c)(d): Gap width against planetary mass at location A, B and C. The axis of the plots are in arbitrary units and do not have the same scale.}
  	\label{fig:app_gpillustration}
\end{figure}
In this section, we schematically illustrate the relation between observed gap width $ \Delta_{\rm gap} $ and planetary mass $ M_{\rm p} $ at three different locations of a disk, and show how the relation corresponds to the interpretations of the different gap groups. As shown by Figure \ref{fig:app_gpillustration} (a), we choose three different $ h_{\rm gap}/R_{\rm gap} $ values that correspond to three fixed locations A, B and C on a disk. At each location, we consider how $ \Delta_{\rm gap} $ varies with the $ M_{\rm p} $, as plotted in Figure \ref{fig:app_gpillustration} (b), (c) and (d). 

According to the planetary formation scenario that we assumed, if $ M_{\rm p} $ is smaller than $ M_{\rm iso} $, the gap is considered to be dust-only gap and $ M_{\rm p} = M_{\rm p,dust}\propto \Delta_{\rm gap}^3 $ (red solid line) given by equation~(\ref{eqn:pltmass_Hill}); otherwise, the gap is considered to open in both the gas and dust disks, therefore $ M_{\rm p} = M_{\rm p,gas} \propto \Delta_{\rm gap}^2 $ (green solid line) as described by equation~(\ref{eqn:crit_gas_width}). The uncertainty of the predicted mass is also taking into account: the red dashed line extends to the upper limit of the mass prediction if the gap is in the dust gap regime, and the green dashed line extends to the lower limit of the mass prediction if the gap is in the gas gap regime. From Figure \ref{fig:app_gpillustration} (b), (c) and (d), the relative positions of the red line and green line are different at different disk locations.

In case A, the projections of the red and green lines onto the $ y $-axis do not span the whole range of $ \Delta_{\rm gap} $. Therefore, any gap width within the grey region does not correspond to any planetary mass. In other words, if a gap is observed at such a width, we can exclude it from the planetary origin, which implies the interpretation of the group IV. This conclusion is the direct consequence if we apply our mass prediction formulae and adopt the planetary formation picture in section \ref{sec:planet_formation_picture}. It is possible that this zero-th order discontinuity of the $ \Delta_{\rm gap} $-$ M_{\rm p} $ relation is not physical but simply due to our poor knowledge of how the gap width changes around $ M_{\rm iso} $. In reality, the sudden shift around $ M_{\rm iso} $ may be replaced by a continuous curve that smoothly connects two regimes. In this case, the planetary interpretation of the gap is not excluded.

In case B, the projections of the red and green lines onto the $ y $-axis span the entire range of $ y$-axis without any overlapping. This case contains no degeneracy, as an arbitrary gap width can always be uniquely explained by a planetary mass, and thus there is no ambiguity about how to interpret the nature of the gap. 

In case C, although the projections of the red and green lines onto the $ y $-axis together cover the entire range of $ y $-axis, there is an overlap that corresponds to the indistinguishable region. If we observe a gap whose width falls into this region, there are two possible mass predictions corresponding to dust gap alone interpretation and gas gap interpretation respectively. Gaps in this indistinguishable region are classified as group III gaps.

\section{Classification of gaps at $ \alpha =\num{e-2} $ and $ \num{e-4} $}
\label{sec:appendix_class_gaps_appendix}
\begin{figure}[h!]
	\centering
	\gridline{\fig{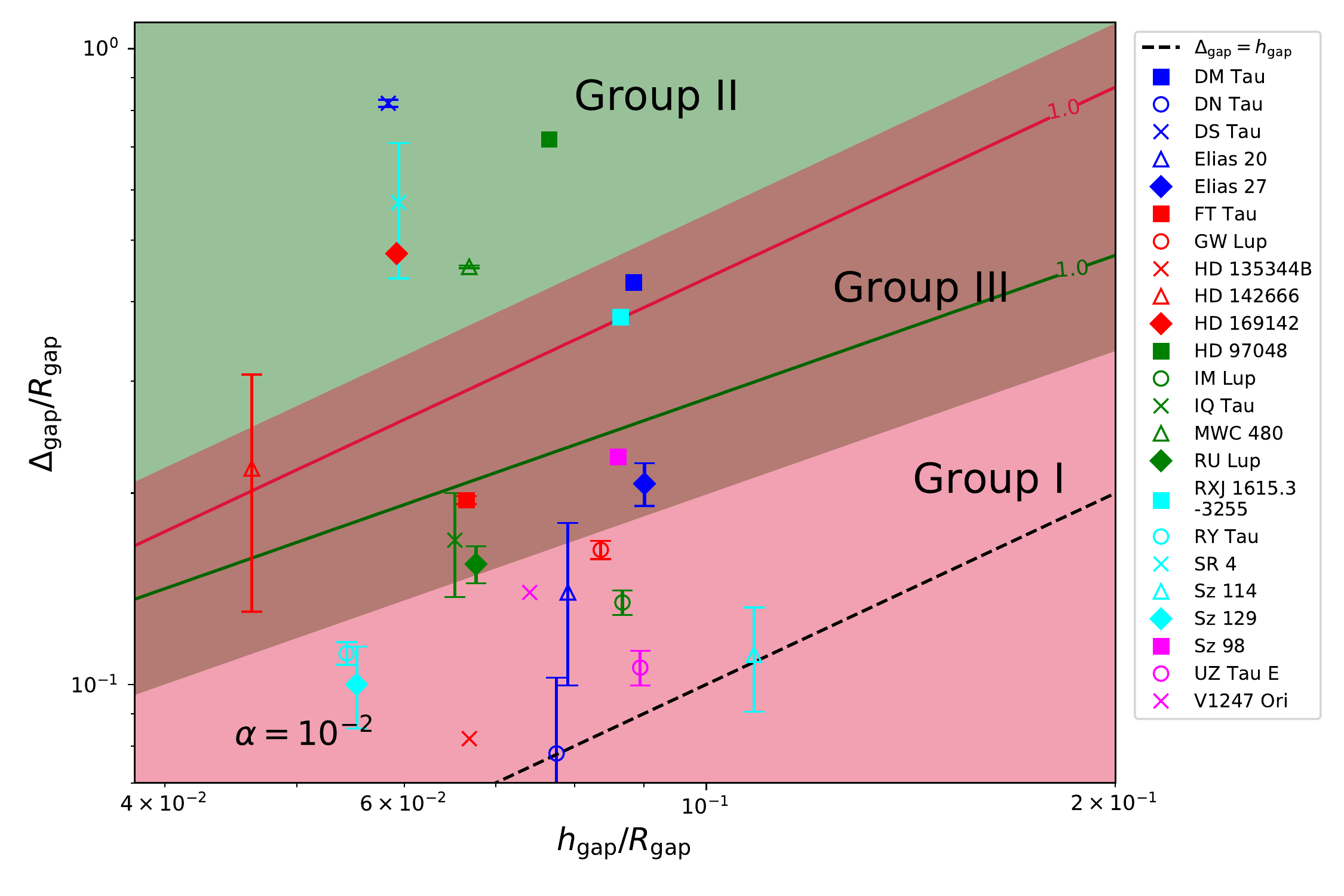}{0.5\linewidth}{(a)}
		\fig{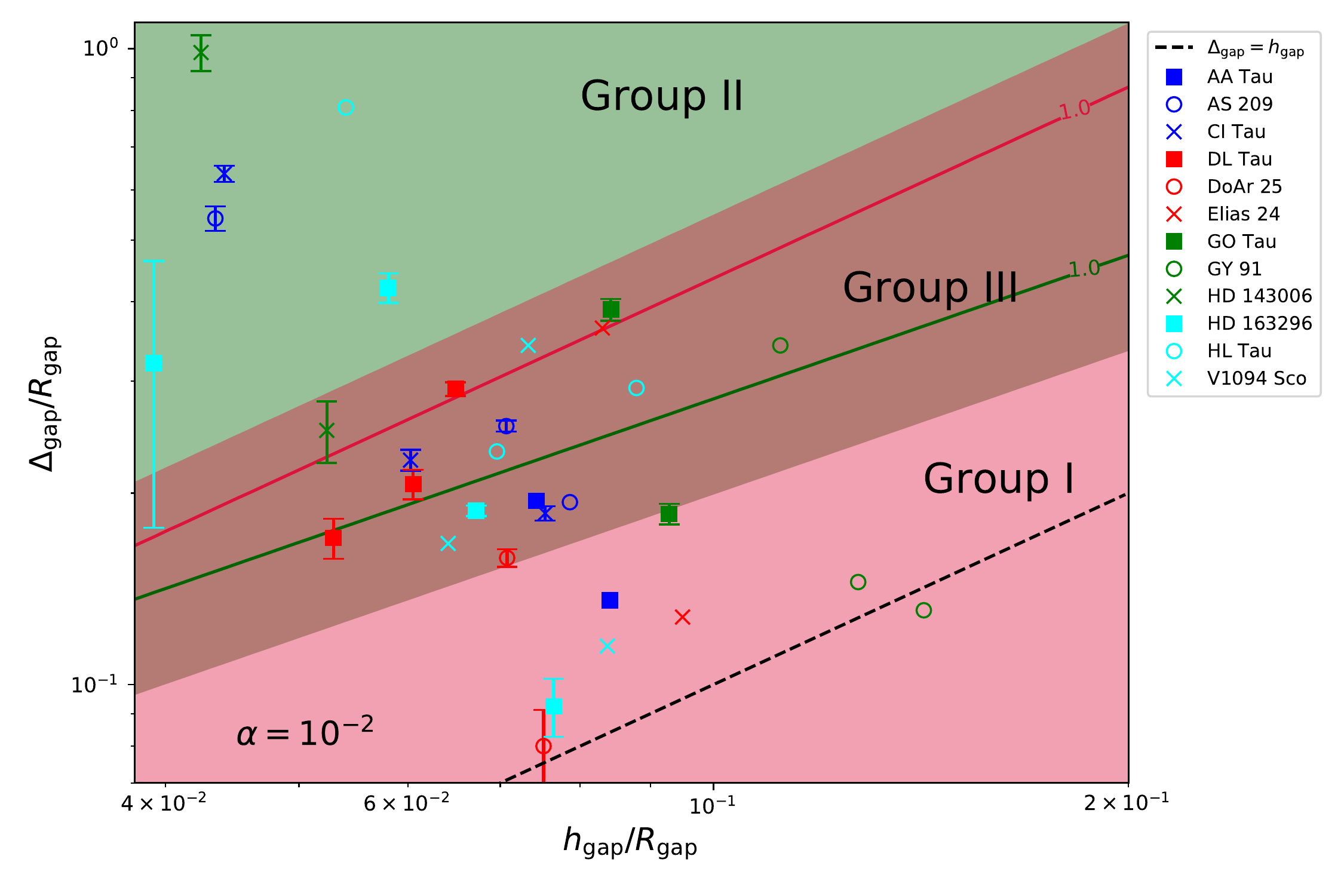}{0.5\linewidth}{(b)}}
	\gridline{\fig{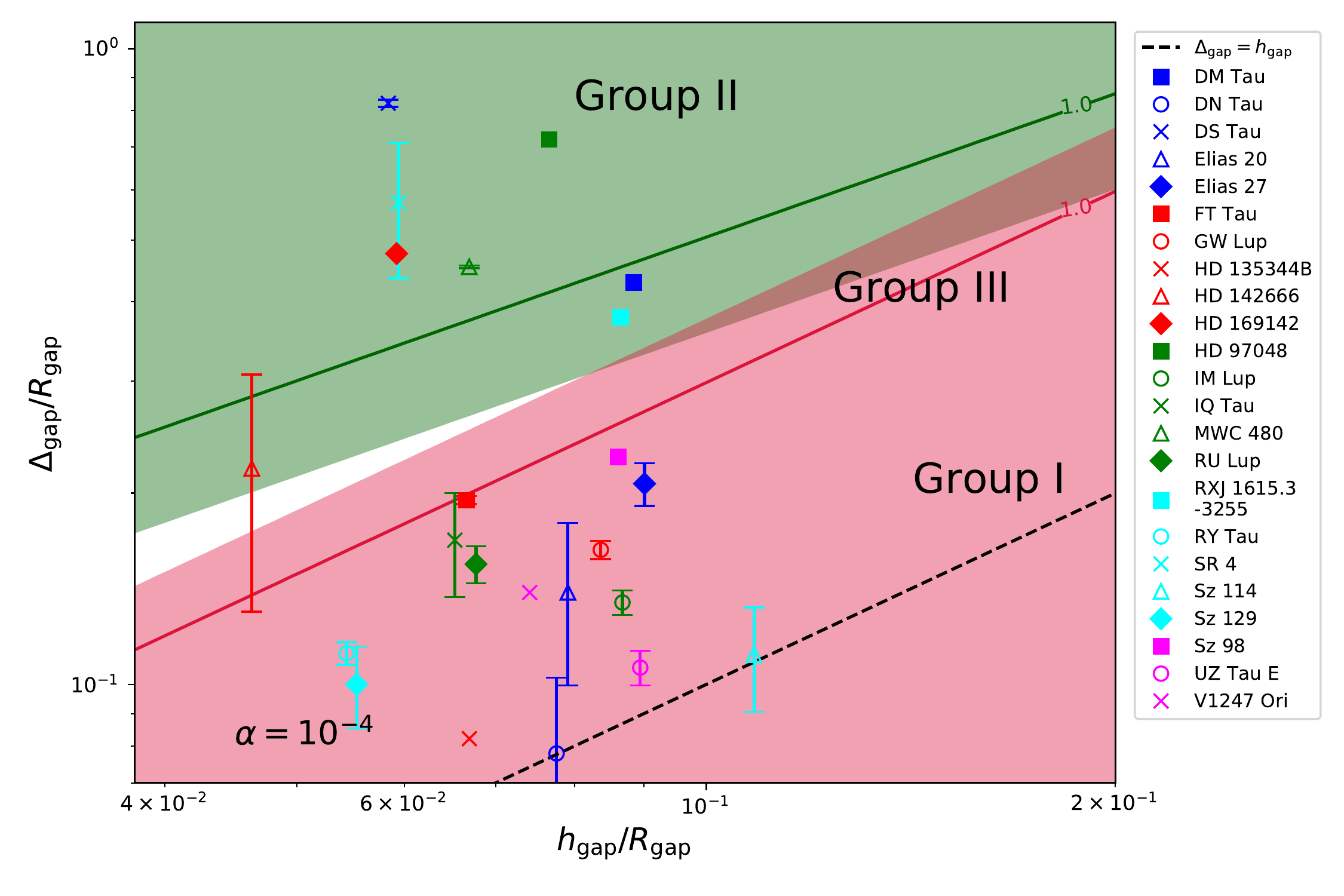}{0.5\linewidth}{(c)}
	\fig{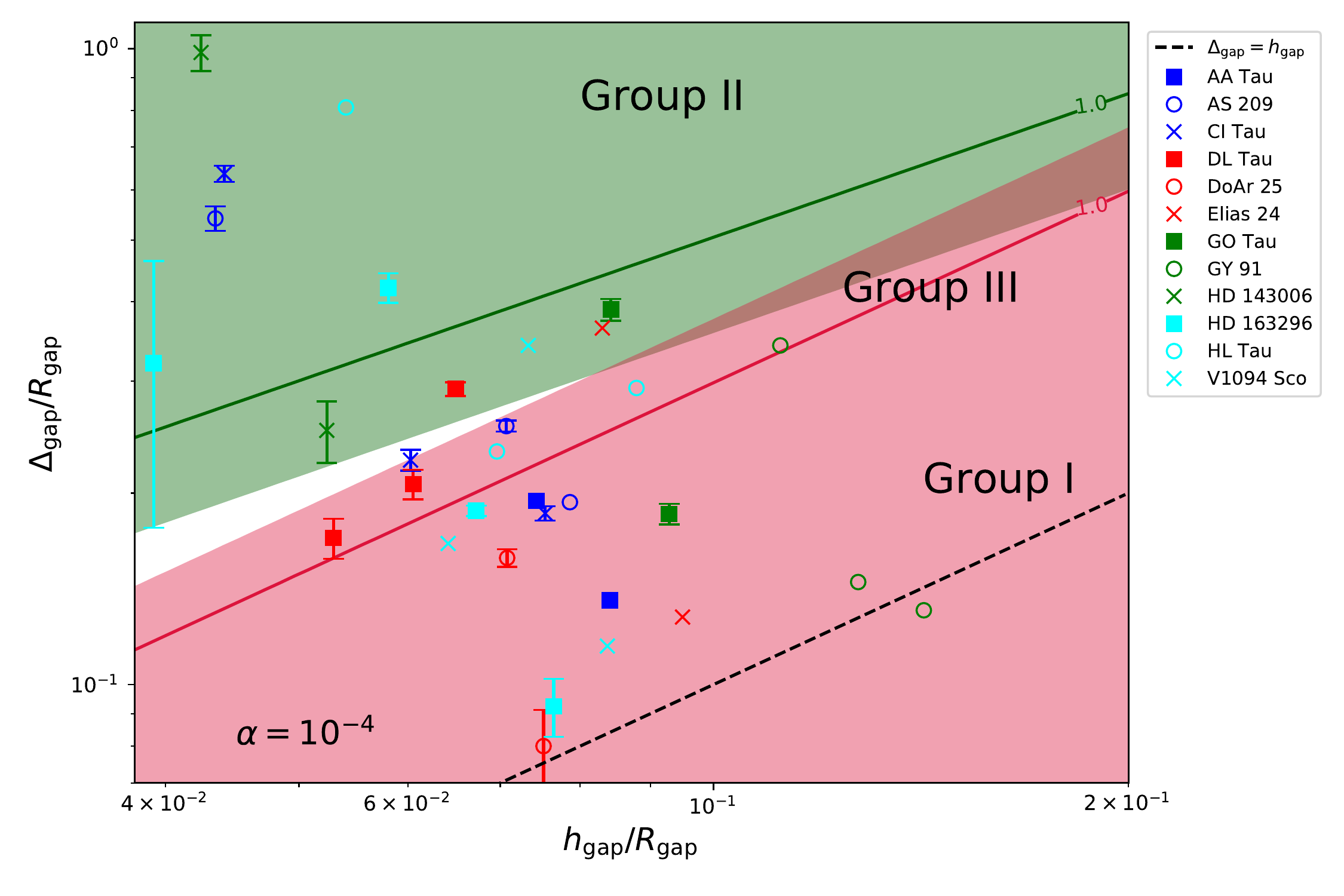}{0.5\linewidth}{d)}}
	\caption{(a)(b): Same as Figure \ref{fig:gap_classification_1e-3} but with $ \alpha = \num{e-2} $. (c)(d): Same as Figure \ref{fig:gap_classification_1e-3} but with $ \alpha = \num{e-4} $.} 
	\label{fig:class_gaps_appendix}
\end{figure}
\clearpage


\end{document}